\documentclass[11pt]{article}
\usepackage{amssymb,amsmath,epsfig}
\usepackage{graphicx}
\usepackage{mathrsfs}
\usepackage{color}
\usepackage[normalem]{ulem}
\usepackage{cancel}
\newcommand{\bbZ}{{\mathbb Z}}
\newcommand{\bbR}{{\mathbb R}}
\newcommand{\bbN}{{\mathbb N}}

\newcommand{\E}[1]{\mathbb{E}\left[#1\right]}

\begin{document}

\title{Statistical Challenges in Microrheology
\thanks{The first author was supported in part by the Louisiana Board of Regents award LEQSF(2008-11)-RD-A-23.  The third author was supported by the NSF grant DMS-1100281, the Cystic Fibrosis Foundation (HILL08I0), and the NIH (R01-HL077546-01A2).  The fourth author was supported in part by the NSF grant DMS-0714939. All of the authors would like to thank Gregory Forest.}
 }

\author{Gustavo Didier \footnote{Mathematics Department, Tulane University.} \and Scott A. McKinley \footnote{Mathematics Department, University of Florida.} \and David B.\ Hill
\footnote{Cystic Fibrosis Pulmonary Research and Treatment Center, University of North Carolina at Chapel Hill.}
\and John Fricks
\footnote{Department of Statistics, Pennsylvania State University.}
\footnote{Correspondence to: John Fricks, Department of Statistics, 326 Thomas Building, Penn State University,
University Park, PA 16802-2111. E-mail: fricks@stat.psu.edu.}}

\date{}
\maketitle

\begin{abstract}
Microrheology is the study of the properties of a complex fluid through the diffusion dynamics of small particles, typically latex beads, moving through that material. Currently, it is the dominant technique in the study of the physical properties of biological fluids, of the material properties of membranes or the cytoplasm of cells, or of the entire cell.  The theoretical underpinning of microrheology was given in Mason \& Weitz (Physical Review Letters; 1995), who introduced a framework for the use of path data of diffusing particles to infer viscoelastic properties of its fluid environment. The multi-particle tracking techniques that were subsequently developed have presented numerous challenges for experimentalists and theoreticians. This paper describes some specific challenges that await the attention of statisticians and applied probabilists. We describe relevant aspects of the physical theory, current inferential efforts and simulation aspects of a central model for the dynamics of nano-scale particles in viscoelastic fluids, the generalized Langevin equation.
\end{abstract}

\section{Introduction}

Broadly speaking, rheology is the study of the flow and deformation of soft matter. Classically, a primary goal of rheologists is to develop constitutive laws for colloids, gels and non-Newtonian fluids. The field is filled with amusing household examples of mathematically exotic behavior: shear-thinning (for example, the way one can help ketchup pour out of a bottle by shearing it with a knife), rod climbing (the way cake batter rises up a stirrer as it is mixed), and shear-thickening (the way some fluids, such as the children's toy Oobleck, exhibit properties of a solid when subjected to shear). Many biological fluids, such as blood, mucus and the cytoplasm of cells, exhibit non-Newtonian properties. Models of diffusion in biological fluids have been developed for pharmaceutical~\cite{2005-suh,suk2009penetration,lai2009mucus} and medical \cite{Dixit:2008} use. These applications exploit both local properties, such as polymer mesh size~\cite{dawson2003enhanced} and viscoelasticity~\cite{guigas2007probing,gardel2006stress}, and global properties, such as heterogeneity of both viscosity and layer thickness~\cite{lai2009altering}.

The rheological study of such biological materials poses a unique practical problem though. Whereas classical rheometers can probe materials by applying both local and global shears and stresses, biological materials tend not to be available in sufficient quantities to apply traditional tests. (That is to say, experimenters find it very difficult to come across liters of mucus on demand.) To confront this constraint, techniques have been developed to probe media locally by embedding and tracking manufactured microparticles~\cite{squires2010fluid}. Dragging a magnetized microbead through a medium is considered \emph{active} microrheology, while simply placing a bead and recording its diffusive motion is considered \emph{passive}. For biological materials, the latter is often preferred because the active process may destroy the medium as it is being measured.

At first blush, it seems unlikely that passive observation of diffusing particles can reveal much about a fluid's microstructure. However, an important step in this direction appeared in the 1995 work {\it Optical Measurements of Frequency-Dependent Linear Viscoelastic Moduli of Complex Fluids}, by Mason \& Weitz (MW). In that work, the authors introduced a theoretical framework for how to use microparticle path data to infer viscoelastic properties of its fluid environment. The multi-particle tracking techniques that were subsequently developed have presented numerous challenges for experimentalists and theoreticians.  In particular, the path of a single particle follows a stochastic process; this contrasts traditional rheological techniques to study non-Newtonian fluids, which are essentially deterministic phenomena.  Our goal is to indicate some specific challenges arising from these stochastically driven experiments that await the attention of statisticians and applied probabilists. We will focus in particular on the background and current practices for inferring what is known as a viscoelastic fluid's \emph{complex viscosity}.  Where possible we will also indicate long-term challenges that exist for other local properties, such as directly characterizing a fluid's microstructure, as well as global properties like quantifying spatial heterogeneity.

\begin{figure}
	\begin{center}
	\includegraphics[scale=0.7]{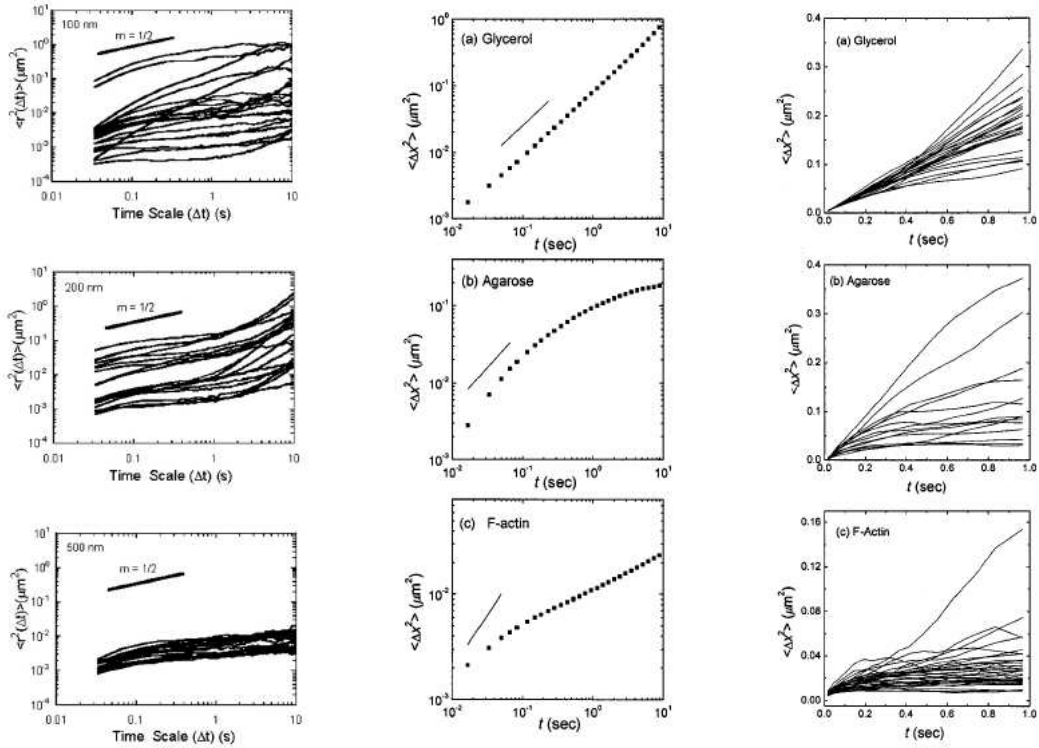}
	\caption{\label{f:anomalous-msds} At left, a snapshot from Dawson et al \cite{dawson2003enhanced} featuring ``pathwise MSDs'' for particles of sizes 100 nm, 200 nm and 500 nm respectively diffusing in sputum from a human with cystic fibrosis. In the center and right are figures from Valentine et al \cite{valentine2001} displaying the ensemble MSD (center) and a collection of pathwise MSDs (right) of microparticles in glycerol, agarose and F-actin, respectively. These data are presented by microrheologists to demonstrate that complex fluids are viscoelastic (mucus, agarose and F-actin) and heterogeneous (mucus and agarose).
}
	\end{center}
\end{figure}

In Section 2, we summarize the MW approach.  The method relies heavily on the theoretical mean-squared displacement (MSD), $\E{X^2(t)}$, of embedded particles. Figure \ref{f:anomalous-msds}, from \cite{2005-suh}, is representative of an important group of papers~\cite{lai2009altering,matsui2006physical,lieleg2010characterization,valentine2001} that rely on MSD concepts.  The authors display an ensemble of what are called \emph{pathwise} or \emph{particle} MSDs, which are notably nonlinear.  The pathwise MSD is essentially a method of moments estimator for the theoretical MSD (see equation (\ref{eq:pathwise-msd}) for a definition). The phenomena of non-linear MSD is referred to as \emph{anomalous diffusion}. There are variety of theoretical hypotheses for what can cause anomalous diffusion. The most prominent explanations include the presence of obstacles, caging, binding, memory \cite{saxton1996anomalous,saxton1994anomalous,levine2000one}.  All of these postulates result in a particle's behavior having memory, and so the MW approach, which coarse-grains these microstructure concerns, is quite general.  They use the \emph{generalized Langevin equation} (GLE) as a universal model.

In the present work, we will follow MW and use the GLE as a starting point. In Section \ref{physics}, we show conditions under which the GLE supports both diffusive and subdiffusive behavior. In addition, we will discuss the physical motivations for the experimental data that is being collected and how this data should connect to specific stochastic models. In Section \ref{inference} we note that the pathwise MSD techniques are subject to substantial variability, which is not well-characterized.  This approach may not be the most efficient method for extracting information about the underlying model.  This is presently overcome by \emph{ad hoc} methods, but in Section \ref{inference}  we discuss the advantages of using spectral methods such as a local Whittle estimator.  First, the estimator provides a useful rigorous test for negating a null hypothesis that a particle path is simply Brownian motion or other purely diffusive process.  Second, since the local Whittle estimator is both more efficient and robust for detecting subdiffusivity, it can be used to quantify heterogeneity among particle paths.  Finally, in Section \ref{simulation}, we address simulation methods including a presentation of a wavelet-based algorithm for the generation of fractional Brownian motion driven GLE.

\section{Physical Models}
\label{physics}

\subsection{Passive microrheology and Newtonian fluids}

For Newtonian fluids such as water, the rheological property of interest is the viscosity $\eta$, and it is a classical observation that the viscosity of a fluid can be inferred by measuring the MSD of suspended diffusing particles. In this section, we will develop this argument with an eye towards the follow-up Section \ref{sec:non-newton}, in which we describe how these notions generalize to non-Newtonian fluids.

Whereas mathematicians and statisticians tend to think of Brownian motion in terms of its statistical properties -- stationary, independent and normally distributed increments with variance proportional to the length of the time increment -- physicists and engineers often employ a Newtonian framework.  The construction of the celebrated Langevin equation relies on the balance of forces ($F = ma$):
\begin{equation}
	m \ddot X = F_{\text{friction}} + F_{\text{external}} + F_{\text{thermal}},
\end{equation}
where $X$ is the position of the particle and $m$ is its mass. For a spherical particle in a viscous fluid, the force due to friction is given by the Stokes drag law, $F_{\text{friction}} = 6 \pi r \eta \dot X$, where $r$ is the radius of the particle and $\eta$ is the viscosity of the fluid.  The external force is often expressed as the gradient of some potential function, $F_{\text{external}} = -\nabla \Phi(X)$.  In a laboratory setting, one might use a quadratic potential to account for the influence of an optical trap \cite{Coppin:1996}, for example.  In polymer physics and molecular dynamics simulations, which involve many particles, it is assumed that there is an intrinsic configuration potential that encodes the energy due to bond lengths and bond angles among the particles in the system \cite{rubinstein2003polymer}.

The remaining thermal force term is meant to summarize the influence of the successive impacts of fluid molecules on the submerged particle. The impacts are assumed to be independent. Because they are frequent it is expected that a central limit theorem will hold and therefore the sum of their effects should be Gaussian.  Furthermore, the variance of the sum should be proportional to both the temperature $T$ of the system and the viscosity of the fluid $\eta$, which both govern the velocity of the fluid molecules and the frequency of impact with the foreign particle. This observation that the variance of the fluctuations and the magnitude of the dissipative force of friction should should be proportional to each other, because they are both caused by the same physical entities in the fluid molecules, dates back to Einstein and is called the \emph{Fluctuation-Dissipation Theorem}.

In modern mathematical notation, \emph{physical Brownian motion} $X(t)$ in one dimension is written as the integral of the velocity process $V(t)$, which is itself the solution to a stochastic differential equation (SDE),
\begin{align}
	\frac{d}{dt} X(t) = V(t), \qquad
	\label{eq:le} m d V(t) = - \gamma V(t) dt - \nabla \Phi(X(t)) dt + \sqrt{2 k_B T \gamma} dW(t),
\end{align}
where $k_B$ is Boltzmann's constant. Often the SDE for the velocity process is known itself as the \emph{Langevin equation}.

For this section, we will restrict our attention to the free diffusion case, when $\Phi(x) \equiv 0$, yielding an Ornstein-Uhlenbeck process for $V(t)$. Also, we will make calculations for diffusion in one dimension, but the calculations easily generalize. We take for the initial condition $X(0) = 0$ and $V(0) = v$ where the random variable $v$ is drawn from the stationary distribution of $V(t)$, $N(0,k_B T / \gamma m)$. One can write an exact solution of the free velocity equation in Duhamel form,
\begin{equation*}
	V(t) = e^{-\frac{\gamma}{m}t} v + \frac{\sqrt{2 k_B T \gamma}}{m}\int_0^t e^{-\frac{\gamma}{m}(t-s)} dW(s),
\end{equation*}
but more important is the autocorrelation function $\rho(t) := \E{V(t_0 + t)V(t_0)}$. Since the initial condition is chosen from the stationary distribution, the auto-correlation does not depend on $t_0$ and is given by
\begin{equation*}
	\rho(t) = \frac{k_B T}{m} e^{- \frac{\gamma}{m}t}.
\end{equation*}

Meanwhile the position process satisfies $\E{X(t)} = 0$. Furthermore,  employing Fubini's theorem and linearity of expectation, we have a relationship between the MSD and the velocity autocorrelation function
\begin{align} \label{eq:msd-rho}
	\E{X^2(t)} &= \E{\int_0^t V(t') dt' \int_0^t V(s') ds'} = 2 \int_0^t \int_0^{t'} \rho(t'-s') dt' ds'.
\end{align}
This simplifies to
\begin{align} \label{eq:msd-x-newton}
	\E{X^2(t)} = \frac{2 k_B T}{m} \int_0^t \frac{m}{\gamma} (1 - e^{-\frac{\gamma}{m}t'}) dt' = \frac{2 k_B T}{\gamma} \Big(t + \frac{m}{\gamma}(1- e^{-\frac{\gamma}{m}t})\Big).
\end{align}

If $\{X_n(t)\}_{t \geq 0}$, $n \in \{1, \ldots, N\}$, are a collection of independent physical Brownian motions as would be the case for a Newtonian fluid assuming the particles were sufficiently spaced as to not interact with one another, then it follows that the empirical ensemble-averaged MSD
\begin{equation} \label{eq:ensemble-msd}
	\langle X^2(t) \rangle := \frac{1}{N} \sum_{n = 1}^N (X_n(t)-X_n(0))^2
\end{equation}
will satisfy
\begin{equation}
	\label{eq:stokes-einstein}
	\lim_{t \to \infty} \frac{\langle X^2(t) \rangle}{t} = \frac{ k_B T}{3 \pi r \eta},
\end{equation}
where we have written the drag $\gamma$ in terms of the viscosity of the fluid $\eta$. Because one can control for the particle radius $r$ by independent means in a laboratory setting, particle tracking can give a robust means to measure a fluid's viscosity. The equation \eqref{eq:stokes-einstein} is known as the Stokes-Einstein Relationship and is crucial to the physical basis for passive microrheology.

\subsection{Passive microrheology and non-Newtonian fluids}
\label{sec:non-newton}

A hallmark property of non-Newtonian fluids is a nonlinear response to applied stress. A shear-thickening material like Oobleck, corn starch suspended in water, will exhibit a super-linear response to increasing applied shear force, while shear-thinning materials, like ketchup or peanut butter, will initially resist like a solid, but will eventually dramatically yield when sufficient force is applied.

\begin{figure}
	\begin{center}
	\includegraphics[scale=0.5]{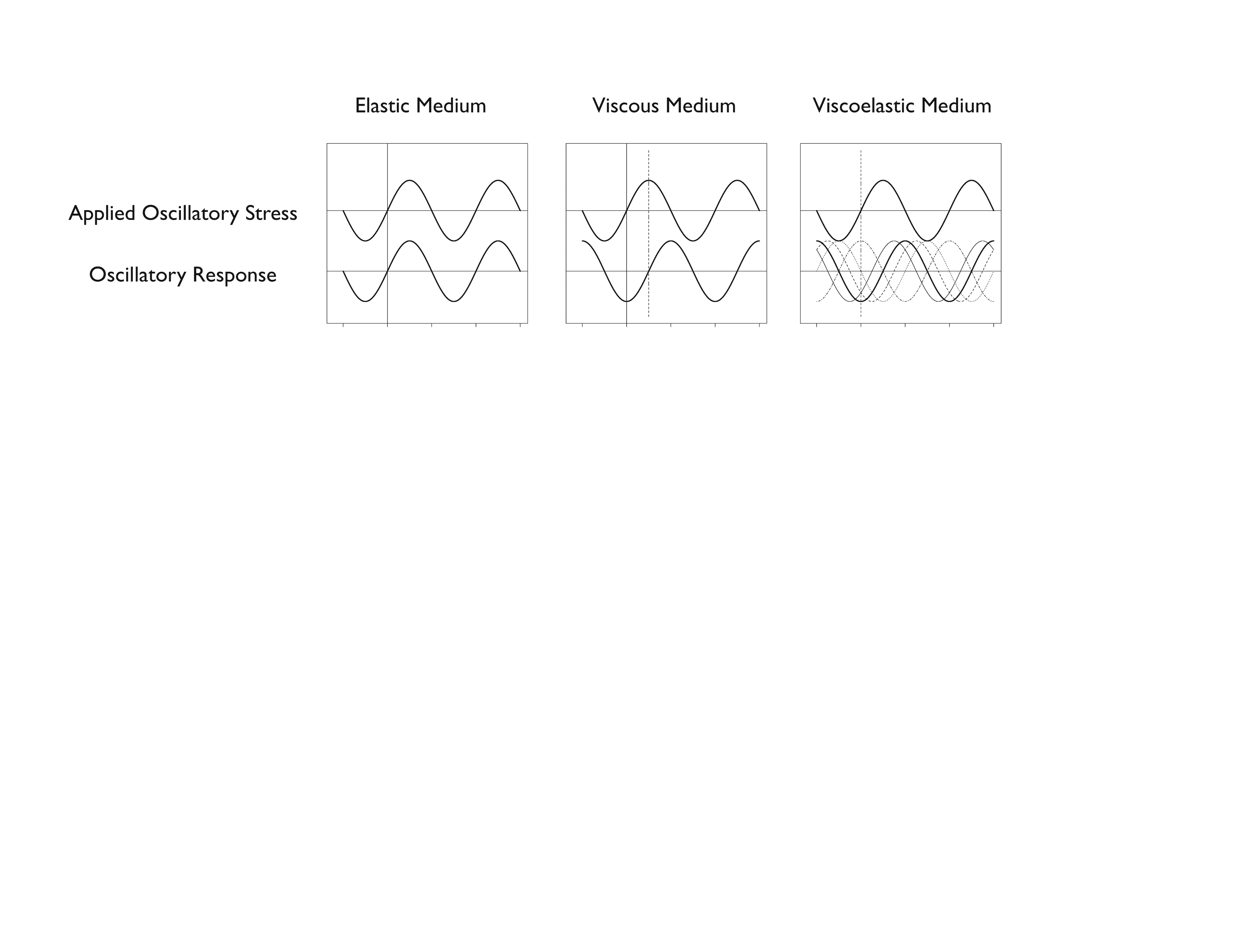} 		\caption{\label{f:shear-modulus} A cartoon depiction of the response by viscous, elastic and viscoelastic fluids to oscillatory stress.  The elastic response is in phase with the stimulus, while the viscous response is out of phase by $\pi/2$ radians. The viscoelastic response is characterized by a superposition of phase-shifted sinusoidal curves. The respective magnitudes of the elastic and viscous responses are often expressed as the storage and loss moduli $G'$ and $G''$ as displayed in Equation \eqref{eq:viscoelastic-response} and Figure \ref{f:masonweitz}.
}
	\end{center}
\end{figure}

A standard procedure for capturing this nonlinear response is to apply a very small oscillatory shear stress. The applied force must be small because dramatic deformation events cannot yet be rigorously characterized. However a sufficiently small applied sinusoidal stress with magnitude $\gamma_0$ and angular frequency $\omega$ will induce an oscillatory response of the form  \cite{larson2001structure}
\begin{equation} \label{eq:viscoelastic-response}
	\theta(t) = \gamma_0 [G'(\omega) \sin(\omega t) + G''(\omega) \cos(\omega t)].
\end{equation}
The functions $G'$ and $G''$ are respectively known as the \emph{storage} and \emph{loss} moduli.  These are so-named because they capture the elastic and viscous components of the material in the following sense.  If a material is purely elastic, the induced stress will oscillate perfectly in phase with the applied force.  On the other hand, a purely viscous fluid will oscillate $\pi/2$ radians out of phase. See Figure \ref{f:shear-modulus} for a visual representation of this phenomenon.

The complex function $G^*(\omega) = G'(\omega) + i G''(\omega)$ is referred to as the \emph{complex shear} modulus and it is directly related to the natural generalization of the Newtonian viscosity, namely, the \emph{complex viscosity} $\eta^*(\omega)$. This relationship is given by\footnote{It is an idiom of the rheology community to use an asterisk to denote a Fourier transformed function.  Because the notation is so entrenched we use $G^*$ and $\eta^*$ but for other instances of the Fourier transform we use a circumflex, $\hat{f}$.} \cite{squires2010fluid}
\begin{equation*}
	G^*(\omega) = i \omega \eta^*(\omega).
\end{equation*}
It is at this point that we rejoin the story of microrheology.

\begin{figure}
	\begin{center}
	\includegraphics[scale=0.5]{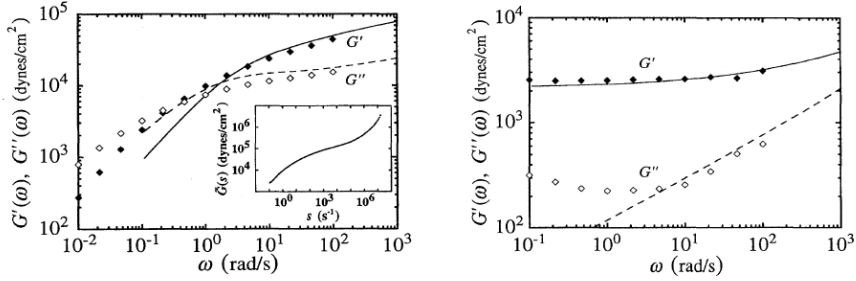}
	\caption{ \label{f:masonweitz} Complex storage and loss moduli reported in the original paper by MW \cite{mason:weitz:1995}. When $G'$ is larger than $G''$, the material acts more like a solid than a fluid.  When the sizes are reversed the material is more fluid-like. For complex fluids, whether the material is more solid or fluid can change depending on the frequency of applied stress. Note that the behavior of these moduli can either approach zero or approach a nonzero constant as the frequency $\omega$ tends to zero. This rheological characterization can be inferred from the asymptotic MSD of embedded microparticles. This connection is laid out in Section 3, equations \eqref{e:asympt_exp}, \eqref{e:alpha=1-2d} and \eqref{e:G-near-zero}.}
	\end{center}
\end{figure}

The fundamental insight provided by  MW is that the natural Generalized Stokes-Einstein Relation (GSER) should be written in the Laplace domain. Denoting the Laplace transform by $\mathscr{L}[f(\cdot)](z) = \tilde f(z) := \int_0^\infty e^{-zt} f(t) dt$ and solving for the behavior of the MSD, equation (\ref{eq:stokes-einstein}) becomes $\mathscr{L}\{\langle X^2\rangle\}(z) = k_B T / 3 \pi r z^2 \eta$. Therefore the hypothesized form for the GSER is \cite{mason:weitz:1995}
\begin{equation}
\label{eq:gser}
	\mathscr{L}\left\{\langle X^2 \rangle\right\}(z) \approx \frac{k_B T}{z^2 (\pi r \tilde \eta(z))}.
\end{equation}
In the next section, we will offer a slightly revised version of MW's development for this form of the GSER. It follows from postulating that the Generalized Langevin Equation (see next section), first proposed by Mori and Zwanzig for the motion of particles in a heat bath~\cite{zwanzig2001nonequilibrium}, is an appropriate model for diffusion in a viscoelastic environment. At the heart of this argument is a connection between the complex viscosity and an assumed memory kernel experienced by diffusing microparticles.

Before moving on, though, we wish to emphasize a detail about the use of the approximation notation in \eqref{eq:gser} rather than equality (compare to similar statements in \cite{mason:weitz:1995} and \cite{squires2010fluid}). Most importantly, mass has been neglected. From this, two issues arise that have a direct impact on the development of statistical methods.  First, the Stokes-Einstein relationship is itself an asymptotic relationship.  One can view it as the result of taking the mass to zero or as waiting a sufficiently long time for inertial effects to become negligible.  These asymptotic limits are equivalent for viscous diffusion, but they are not necessarily equivalent for viscoelastic diffusion. As was demonstrated in \cite{mckinley:yao:forest:2009}, the zero-mass limit may be singular, creating a term that can be lead to a misinterpretation of any observed transient anomalous diffusion.  Second, as we will show in the next section, any large-time limit will neglect important transient features in the data.  An entire family of classical non-Newtonian fluids known as Maxwell fluids will appear identical in a large-time MSD limit.

\subsection{The Generalized Langevin Equation}\label{sec:GLE}

As is pointed out by Kubo \cite{kubo1966fluctuation}, embedded in the development of the Langevin equation \eqref{eq:le} is an implicit assumption that there is time scale separation between the dynamics of the large foreign particle and those of molecules comprising the surrounding medium. However, in the viscoelastic regime, one cannot assume that the influences of the various fluid molecule impacts are independent of recent activity by the foreign particle.  This manifests itself in two ways: first, this introduces a non-trivial autocorrelation structure in $F_{\text{thermal}}$; second, the friction coefficient should no longer be constant. This non-Markovian property is modeled by way of a memory kernel, which we denote by $\Gamma(t)$, and based on physical considerations, the fluctuation-dissipation theorem implies that the memory kernel should be the same for both thermal fluctuation and friction.  The \emph{generalized Langevin equation} (GLE) for the velocity process is therefore written
\begin{equation} \label{eq:gle}
	m\frac{d}{dt} V(t) = -\gamma \int_{-\infty}^t \Gamma(t-s) V(s) ds - \nabla \Phi(X(t)) + F(t),
\end{equation}
where $F(t)$ is a stationary Gaussian process with autocorrelation function
\begin{equation} \label{eq:gle-acf}
	\E{F(t)F(s)} = k_B T \gamma \Gamma(|t-s|).
\end{equation}
The lower limit of integration is $-\infty$ to emphasize that the velocity process is stationary. In this section we will continue to take the potential $\Phi(x) \equiv 0$ for expository purposes, and we will only consider 1-d dynamics.  It is generally assumed that the velocity processes corresponding to each coordinate of a diffusing particle are independent.

We redefine the drag parameter to be $\gamma = 6 \pi r \eta_0$, where $\eta_0$ is the viscosity of the fluid when there is no applied stress. The crucial physical assumption then is that the complex viscosity $\eta^*$ is related to the Fourier transform of the memory kernel via the relation $\hat \Gamma(\omega) = \eta^*(\omega) / \eta_0$. For example, because the complex viscosity is constant for purely viscous fluids, the memory kernel should be a Dirac $\delta$-distribution.  Indeed, the GLE reduces to the Langevin Equation for this choice of $\Gamma$.

For the simplest viscoelastic fluids, known as Maxwell fluids, $\Gamma$ has the form of a  Prony series:
\begin{equation}
	\label{eq:prony}
	\Gamma(t) =  \sum_{n=1}^N c_n e^{- \lambda_n t}.
\end{equation}
As an example, in the original paper of MW \cite{mason:weitz:1995}, the authors introduce a proof of principle for suspended silica particles in ethylene glycol.  They effectively assumed prior knowledge of four distinct relaxation times (which have the form $\lambda_n^{-1}$) and used the Laplace transform of the empirical MSD to fit the coefficients of the terms.

When the number of modes is extremely large, in a polymer melt for example, such a Prony series can mimic a power law \cite{kou:2008,mckinley:yao:forest:2009}. In this case, it is customary to assume that the thermal fluctuations are given by fractional Gaussian noise (fGn). The latter is the increment process of fractional Brownian motion (fBm), which is the unique Gaussian, self-similar, stationary-increment process (see \cite{embrechts:maejima:2002}). The associated memory kernel has the form
\begin{equation}\label{eq:kernel_fGn}
	\Gamma(t) = 2H(2H-1)|t|^{2H-2}, \quad t \neq 0, \quad \frac{1}{2} < H < 1,
\end{equation}
where $H$, called the Hurst parameter, characterizes the fGn/fBm. The kernel (\ref{eq:kernel_fGn}) gives rise to a qualitatively different regime in terms of the behavior of suspended microparticles, with important consequences for the inferred complex viscosity.  See Sections \ref{s:subdiffusive} and \ref{s:fGLE_integ_repres}.

In general, because $\Gamma$ appears as the covariance of a stochastic process, it is required to be a positive definite function.  By Bochner's theorem, this means that there exists a real-valued positive Borel measure $\hat \Gamma$ such that $\Gamma(t) = \frac{1}{2 \pi} \int_\mathbb{R} e^{- i t \omega} \hat \Gamma(d\omega)$. Deterministic linear integrodifferential equations of a similar type to \eqref{eq:gle} have been well-studied \cite{gripenberg1990volterra}.  If the thermal fluctuation is sufficiently regular\footnote{One can still make sense of the GLE when the thermal fluctuation $F$ is rougher.  For example, when the forcing term is a fractional Gaussian noise there is a singularity at 0 in the memory kernel that results in $F$ only being well-defined in the sense of distributional derivatives.  In Section \ref{simulation} we briefly recount the approach proposed by Kou \cite{kou:2008} where the spirit of the following discussion still holds.}, for example in $C(\mathbb{R})$ or $L^1_{loc}$, then \eqref{eq:gle} has a path-by-path interpretation.  Such solutions can be written in variation of constants form
\begin{equation} \label{eq:gle-v-soln}
	V(t) = \int_{-\infty}^t \chi(t-s) F(s) ds,
\end{equation}
where $\chi(t)$ is known as the \emph{differential resolvent} in the integral equations literature \cite{gripenberg1990volterra} and as the \emph{susceptibility} in the physics literature \cite{adelman1976fokker}. This function $\chi$ is the \emph{fundamental solution} to the homogeneous part of \eqref{eq:gle} in the sense that it satisfies
\begin{equation*}
	m \frac{d}{dt} \chi(t) = -\gamma \int_{-\infty}^t \Gamma(t-s) \chi(s) + \delta (t)
\end{equation*}
Taking the Fourier transform of this equation yields the following frequency-space representation for $\chi$,
\begin{equation*}
	\hat \chi(\omega) = \frac{1}{m i \omega + \gamma \hat \Gamma(\omega)}.
\end{equation*}
To complete our characterization of the velocity process, we note that since the thermal fluctuation term is Gaussian and stationary, the linear transformation \eqref{eq:gle-v-soln} will yield another Gaussian stationary process.
A calculation then reveals that $\rho(t-s) = \E{V(t)V(s)}$ satisfies
\begin{equation} \label{eq:rho-fourier}
	\hat \rho(\omega) = \frac{k_B T \gamma \hat \Gamma(\omega)}{\big|m i \omega + \gamma \hat \Gamma(\omega)\big|^2}.
\end{equation}
This sets the fundamental relationship between the velocity autocorrelation function and the complex viscosity.

In order to derive the GSER \eqref{eq:gser} we observe that the mass of these microparticles is vanishingly small.  Setting the mass equal to zero in the Laplace domain is equivalent to taking a certain weak limit of a sequence of solutions to the GLE with decreasing parameter $m$ that may neglect singular correction terms \cite{mckinley:yao:forest:2009}. If we nevertheless permit ourselves to take this step for expository purposes, \eqref{eq:rho-fourier} reduces to $\hat \rho(\omega) = k_B T/\gamma \hat \Gamma(\omega) = k_B T \eta_0/\gamma \eta^*(\omega)$. On the other hand, the sequence of equalities \eqref{eq:msd-rho} implies that the Laplace transform of the autocovariance is $\tilde \rho(z) = z^2 \mathscr{L}\{\langle X^2 \rangle\}$. Combining these observations yields the GSER \eqref{eq:gser}, which expresses complex viscosity in terms of the MSD.



\subsection{The Generalized Langevin Equation and Anomalous Diffusion}

One of the most frequently reported properties in single particle tracking experiments is that the MSD scales sublinearly over time.  In fact, this is the major qualitative difference between what one expects to see from diffusion in a Maxwell fluid versus diffusion in a polymer melt.

To be precise, we say that a stochastic process $X(t)$ is
	\begin{equation}
		\text{asymptotically }
		\left\{
		\begin{array}{c} \text{subdiffusive} \\
				\text{diffusive} \\
				\text{superdiffusive}
			\end{array} \right\}
				\text{ if } \lim_{t \to \infty} \frac{\E{X^2(t)}}{t} =
					\left\{ \begin{array}{c} 0 \\ \sigma^2 \\ \infty
					\end{array} \right\},
	\end{equation}
where $\sigma$, when finite, is the diffusion coefficient.

Morgado et al \cite{2002-morgado-anom-diff-test} demonstrated that solutions to the GLE can exhibit any of the above behavior depending on the Laplace transform of the memory kernel.  Suppose for example that $X(t)$ is a solution to \eqref{eq:gle} with a memory kernel $\Gamma$ that satisfies $\int_z^\infty \tilde \Gamma(s) / s^2 ds < \infty$ for any $z > 0$. One can show that if
	\begin{equation} \label{eq:diff-test}
		\lim_{z \to 0} \tilde \Gamma(z) =
			\left\{ \begin{array}{c} \infty \\ \sigma^2 \\ 0
			\end{array} \right\},
		\text{ then the process } X(t) \text{ is asymptotically }
		\left\{
		\begin{array}{c} \text{subdiffusive} \\
				\text{diffusive} \\
				\text{superdiffusive}
			\end{array} \right\}.
	\end{equation}
To see this, recall the relationship between the MSD of the position process and the ACF of the velocity process \eqref{eq:msd-rho}:
	\begin{align*}
		\E{X^2(t)} &= 2 \int_0^t \int_0^{t'} \rho(t',s') dt' ds',
	\end{align*}
	where $\rho(t,s) = \E{V(t)V(s)}$. By the final value theorem, it follows from \eqref{eq:msd-rho} that
	\begin{align*}
		\lim_{t \to \infty} \frac{\E{X^2(t)}}{t} = \lim_{z \to 0} z \mathscr{L}\Big\{\frac{\E{X^2(t)}}{t}\Big\}(z) = \lim_{z \to 0} z \int_z^\infty \mathscr{L}\Big\{\E{X^2(t)}\Big\}(s) ds = \lim_{z \to 0} \tilde \rho(z).
	\end{align*}
	The conclusion \eqref{eq:diff-test} then follows from \eqref{eq:rho-fourier}.

This proposition implies that diffusion in Maxwell fluids is asymptotically diffusive, while diffusion in power law fluids will be asymptotically subdiffusive. This dichotomy plays a large role in the discussion of inference in the next section.  In the subdiffusive case, see Section \ref{s:subdiffusive}, an important challenge is in identifying the asymptotic subdiffusive exponent, which in turn dictates the near zero behavior of the complex viscosity.  In the diffusive case, see Section \ref{s:diffusive}, one must capture transient phenomena in order to distinguish among various model fluids.

\subsection{Modeling Challenges}

The preceding discussion has been limited to modeling ensembles of independent particles diffusing in a highly idealized version of biological fluids.  We have neglected tremendously important factors, including surface chemistry of the particles, rigid obstacles in the medium, interaction with boundaries, and heterogeneity in general. For an example of the impact of such factors, we mention recent work by the Hanes lab, who exploit surface chemistry effects for pharmaceutical applications.  By coating particles in polyethylene glycol (PEG), the particles are rendered mucoinert and are observed to move much more quickly \cite{2005-suh,lai2009mucus}. Interestingly, the method reveals certain details concerning the size of gaps in the polymer-mesh that constitutes human mucus.  Because coated 200 nm beads diffuse faster than similarly coated 500 nm beads, this provides evidence that the effective mesh size is somewhere on this scale.

The Hanes lab discovery demonstrates an important principle of passive microrheology.  All observations are properties not of the fluid medium, but of \emph{this} particle in \emph{that} fluid. Therefore, when trying to capture rheological properties of the fluid itself, one must aggregate observations from multiple experimental techniques.  This is one reason why two-point (and $n$-point in general) rheology is vital to the future of the field.  Multiparticle tracking has the capacity to yield information about both local and global properties of a viscoelastic fluid.  At a local level, it is hypothesized that fluid-particle dynamics can vary greatly with distance. For example, in cystic fibrosis mucus studies, it is hypothesized that depending on the surface chemistry of the particle, there may be a depletion layer of the mucosal network in the immediate vicinity of the particle.  This may have a profound effect on whether a substantial number of individuals in an ensemble become immobilized due to binding.  Determining this local environment is impossible with single bead observations though, so several recent efforts have sought to characterize the cross-correlations between beads \cite{levine2000one,hohenegger2008two,hohenegger2008modeling}.
Furthermore, depending on how rigid the fluid structure is, particles may ``talk'' to each other across long distances, as is the case for particles suspended in gels.

Understanding long-range interactions among particles reveals a final inadequacy in the current theory.  Still missing is a comprehensive model for the interaction between a viscoelastic medium and submerged microparticles that has the same level of sophistication that the stochastic immersed boundary method \cite{atzberger2007stochastic} provides for viscous diffusion. A robust model, coupled with accurate and efficient simulation techniques, are a prerequisite for characterizing whether or not variation seen in a given particle ensemble is greater than what should be expected from a homogeneous medium.  We give an example in the next section for how difficult this has proved for simple viscous fluids, revealing how much is left to be done.

\section{Statistical Inference} \label{inference}

\subsection{Pathwise Mean Squared Displacement}

The data displayed in Figure \ref{f:anomalous-msds} is representative of numerous articles in rheological journals. In displaying an ensemble of \emph{pathwise} MSDs (see equation (\ref{eq:pathwise-msd})), the authors are communicating two important features simultaneously: the behavior of a thermally fluctuating microparticle in a viscoelastic medium is plainly distinct from classical diffusion, and furthermore the fluid medium itself can be highly heterogeneous.

Qualitative evidence for heterogeneity can be found in the high variability of the MSD plots of repeated experiments. If one could propose a credible model for viscoelastic diffusion, then the empirical spread of pathwise MSDs can be compared to a simulated distribution.  When the former is significantly larger than the later, this is evidence for fluid heterogeneity.  An early attempt in this direction was performed by Valentine et al~\cite{valentine2001}. In this work, the authors used higher order statistics to compare empirical pathwise MSD curve distributions of microparticles in glycerol, agarose and \emph{F} actin to the spread generated by simulated Brownian motions.


The desire to see heterogeneity compels the use of pathwise statistics to distinguish microrheological paths from classical diffusion. The approach is successful in that often the pathwise MSD curves are clearly sublinear and sometimes appear to go through a series of transitions. One often observes an initial inertial regime where there is ballistic (slope on a log-log plot is two) growth~\cite{fricks:yao:elston:forest:2009,mckinley:yao:forest:2009}.  Over an intermediate time scale subdiffusive (slope less than one) behavior is observed, followed by large time scale behavior that may be subdiffusive, diffusive, or superdiffusive.

As discussed in the previous section, the subdiffusive and diffusive long-time asymptotic behaviors are natural outcomes for the GLE with power law and exponential tail memory kernels, respectively.  While it is true that the GLE also supports super-diffusive behavior, our preliminary analysis of experimental data indicates that ballistic large time behavior results when there is a dominant drift that has not been correctly removed in the statistical analysis.  To this end, it is also possible to introduce artificial \emph{subdiffusive} behavior by removing a mean inappropriately, and so great care must be taken when interpreting the large time regime.

For the purposes of the present discussion, we will assume that in a given time window $I$, we can express what one might call the ``local'' MSD in the following form,
\begin{equation}\label{e:asympt_exp}
\E{X^2(t)} \approx \sigma t^\alpha, \quad \alpha > 0, \quad t \in I.
\end{equation}
The microparticle is said to be diffusive, subdiffusive or superdiffusive if the MSD exponent $\alpha$ is equal to, less than or greater than 1, respectively. This kind of transition behavior for MSD is not unprecedented.  In polymer physics, for example, it has been observed that the MSD of a single bead in a thermal fluctuating bead-spring chain will undergo MSD transitions \cite{rubinstein2003polymer,kremer1990dynamics}.

In single particle tracking techniques, the local MSD is typically estimated by means of a heuristic regression-based method, which we will call \emph{pathwise MSD}. Suppose that a microrheological experiment generates a tracer bead sample path with observations $X(\Delta j)$, $j=0,1, ..., N$. Let $\mu_2(t) := EX^2(t)$. Under (\ref{e:asympt_exp}), assuming the increments are stationary, for $h=1,2,...,n$ and $n << N$, one hopes that
\begin{equation}
	\label{eq:pathwise-msd}
	\bar{\mu}_2( \Delta h  ) := \frac{1}{N-h+1}\sum_{j=0}^{N-h} \left(X(\Delta (j+h) )-X(\Delta j )  \right)^2 \approx  \E{X^2(\Delta h)}, \quad \textnormal{if $N$ is large}.
\end{equation}
One then considers the linear regression specified by
\begin{equation}\label{e:regression}
 \log(\bar{\mu}_2(\Delta h)) = \log(\sigma) + \alpha \log( \Delta h  ) +\varepsilon_h, \quad h=1,...,n,
\end{equation}
and applies ordinary least squares to obtain the estimator $(\widehat{\log}(\sigma), \widehat{\alpha})$.

\begin{figure}
	\begin{center}
	\includegraphics[height=2in]{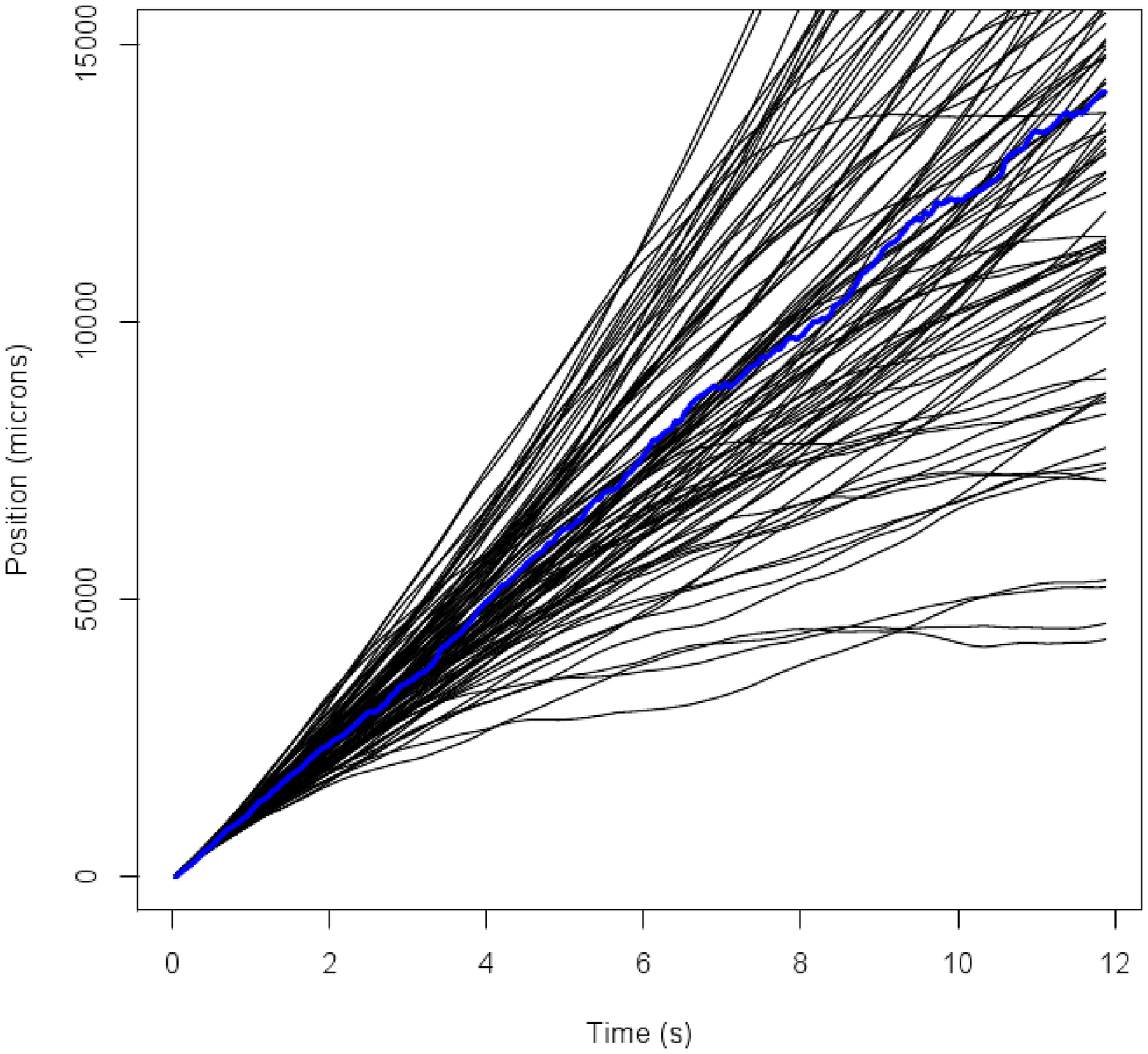} \ 	\includegraphics[height=2in]{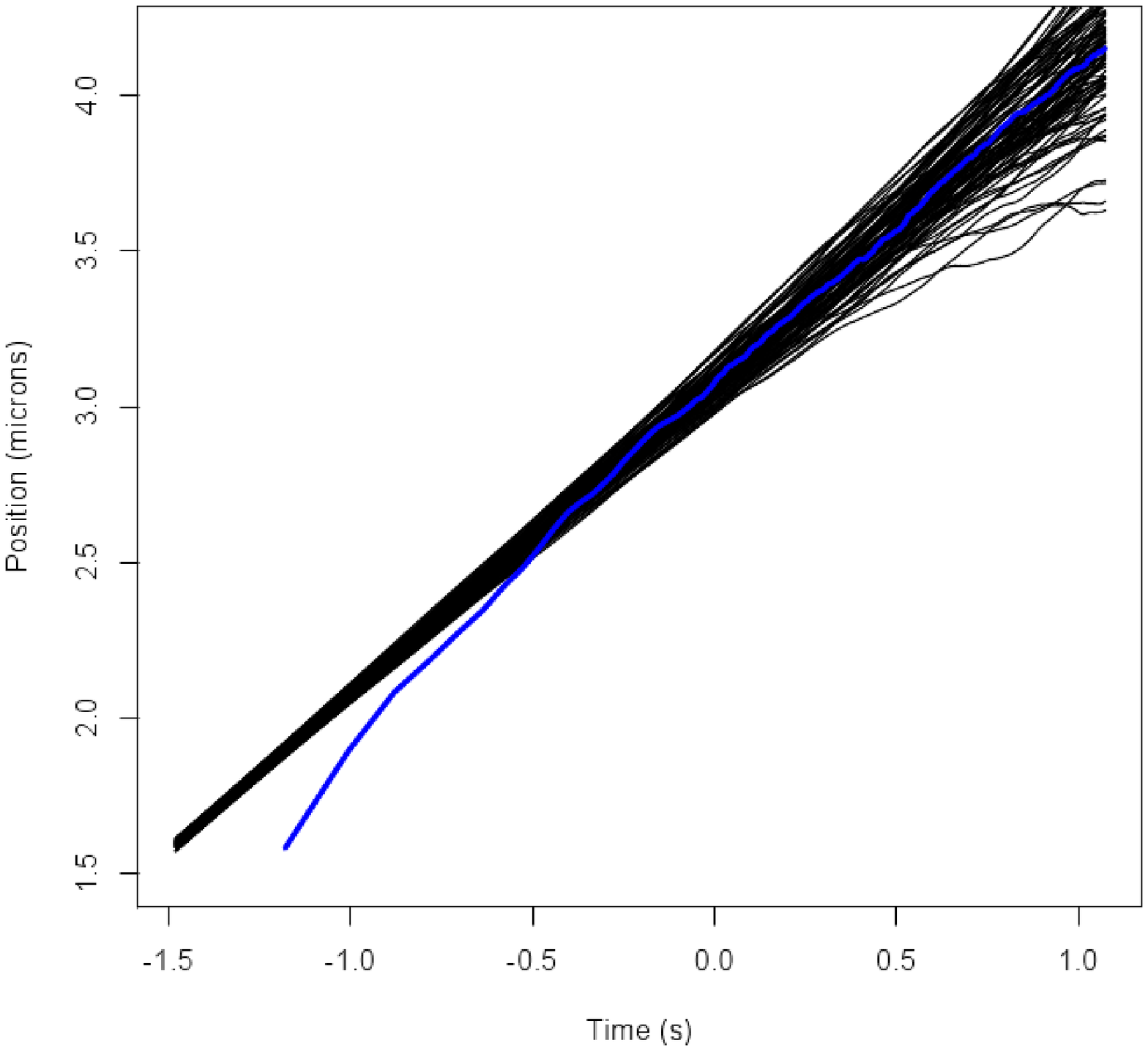} \ %
	\caption{\label{f:water_simulations} At left, a plot of the pathwise MSD for 100 Brownian paths with a diffusivity that corresponds to a 200 nm radius spherical particle in water at $20^0$ C. At right is the same pathwise MSD displayed on a log-log plot, which is prominent in the literature.  Using a log-log plot hides the considerable variation in the empirical estimate for the viscosity of the fluid.  Furthermore, a substantial number of paths will appear subdiffusive when subjected to algorithmic analysis.  (See also Figure \ref{mucusacf}.)
	}
	\end{center}
\end{figure}



However, this procedure is subject to several shortcomings. First, the problem of delineation of regimes is inherently ill-formed, so the methods are naturally \emph{ad hoc}. One must always ``eyeball it'' to decide when the MSD curve looks approximately straight on a log-log plot, and typically the transitions between regimes only become evident when an entire ensemble of pathwise MSDs are superimposed.

As mentioned in the last section, determining whether particle paths are asymptotically diffusive or subdiffusive is a fundamental dichotomy in interpreting experimental results. However, data for large time lags is scarce and when available it may be overwhelmed by spurious drift in the fluid medium. Even after correcting for these physical constraints, the error between pathwise MSD and the true diffusivity of the particle
is itself subject to high variability and induces dependencies as an artifact of the technique.
The problem arises because the error terms $\varepsilon_h$ in (\ref{e:regression}) need to be at least approximately independent and identically distributed for standard regression properties to apply. However, this assumption is violated in the cases of interest. Indeed, as the lag becomes larger, the variance of $\bar{\mu}_2(\Delta h)$ tends to increase. Also, neighboring (and even distant) estimators $\bar{\mu}_2(\Delta h)$ are formed from the same observations, thus implying that they are \textit{dependent} even in the diffusive case. Moreover, the convergence speed of the estimators $\bar{\mu}_2(\Delta h)$, as a function of $N$ can be slow due to the dependence structure of the sequence $ \left(X(\Delta(j+h)-X(\Delta j) \right)^2$, $j=0,...,N-h$, in both diffusive and subdiffusive situations.

The inaccuracy of the pathwise MSD can be at least partially compensated by the use of large amounts of data. However, the number of observations of laboratory generated tracer bead paths is usually limited to no more than 5,000. The left plot in Figure \ref{Brownian_alpha_hat} illustrates the performance of the pathwise MSD based on Monte Carlo runs of Brownian motion sample paths ($\alpha = 1$). Even though the increments of the underlying process are actually independent, the histogram for the distribution of the $\alpha$ as estimated by the pathwise MSD is concentrated in the very wide range $[0.4,1.4]$.

Another serious drawback to the pathwise MSD based methods is that they lack an obvious way to quantify sampling error. For instance, a natural technique would be to perform parametric bootstrap. One assumes that the estimated MSD is the true value and then generates corresponding paths. However, there are multiple memory kernels giving rise to the same theoretical MSD, and thus it is impossible to capture all the stochastic behavior of the underlying stochastic process based on an estimate of $\alpha$ alone. Nevertheless, the ability to quantify
estimation error is especially important when trying to detect heterogeneities across multiple paths in a sample. The inability to quantify sampling variation has been a major impediment to the development of testing procedures for microrheological inferential techniques.


%




\subsection{The Diffusive Case} \label{s:diffusive}

Assuming one can infer that a process is asymptotically diffusive, the focus of statistical investigation should be the transient anomalous behavior. Because the rheological statistic of interest -- the complex viscosity $\eta^*$ -- is essentially the Fourier transform of the memory kernel $\Gamma$, it suffices to directly pursue the memory kernel in the time domain when possible.
Fricks et al~\cite{fricks:yao:elston:forest:2009} developed such a method in the case where the memory kernel is assumed to be a Prony series; recall equation \eqref{eq:prony}. This is the class of memory kernels associated with well-studied Maxwell fluids~\cite{ferry1980viscoelastic}, and therefore any microrheological methodology should be able to reproduce characterizations from classical rheology.

The basic idea of the method is to use a higher dimensional linear stochastic differential equation representation of the generalized Langevin equation, a fact that was previously developed in several forms~\cite{grigolini:1982,mori:1965}.   Since the resulting process is a finite-dimensional linear It\^o stochastic differential equation, the conditional mean and variance can be calculated for the sampled data permitting the use of standard Kalman methods to derive a likelihood function on even coarsely sampled data.

One of the benefits of this Kalman method is that standard likelihood theory applies giving asymptotic estimates of standard errors of the parameters.  Given the likelihood function, a Bayesian method could also be devised allowing for similar quantification of the variability of the estimation scheme.  The ability to quantify estimation error is especially important when trying to detect heterogeneities across multiple paths in a sample.

Kou and Xie \cite{kou:xie:2004} used a version of the GLE to study the subdiffusivity of a particle attached to a protein; however, their methods could be adapted to the diffusive case.  In particular, they used a form of the GLE with a quadratic potential \cite{min2005observation}. By using Kou's representation of the covariance of the resulting equation, they were able to use a method of moments type estimator to fit the Hurst parameter to experimental data \cite{kou:2008}.  One could attempt to modify this method using   a similar representation of the covariance of the increment process for the differenced position data  to apply Bayesian or maximum likelihood methods to the resulting Gaussian process.   Note, however, that this strategy could be computationally very intensive and its success could largely depend on the form of the memory kernel selected.  Both of these are examples of methods to fit microrheological data in a modern statistical context while maintaining fidelity with the well-established physical models, though both use simplifying assumptions to yield a more tractable model.

\subsection{Detecting Subdiffusion } \label{s:subdiffusive}

%

Exploratory investigation of experimental particle path data on any hydrogel, and mucus in particular, indicates that the qualitative behavior of the submerged particles depends strongly on size and surface chemistry. Even among apparently identical particles, significant variation is seen, presumably because
mucus has a highly heterogeneous mesh size distribution that interacts differently with particles of different radii. Typically these properties are reported in terms of highly variable pathwise MSD curves, and in terms of labeling subpopulations of a particle ensemble as \emph{diffusive}, \emph{hindered} or \emph{immobile}. This pathwise classification, which has important consequences in applications to drug delivery, poses a very difficult statistical question: Is it possible to successfully label a given path as being from a diffusive or subdiffusive ensemble, and how much data is necessary to accomplish this task reliably? 

\begin{figure}
	\begin{center}
	\includegraphics[height=2in]{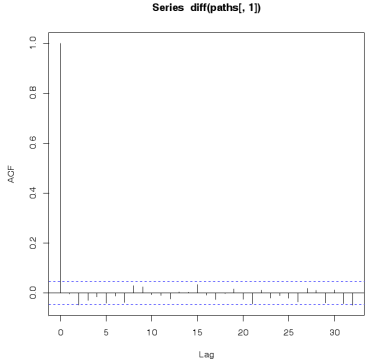}
	\caption{ \label{mucusacf} The ACF of a single path of 1800 observations taken of a particle in human lung mucus with 2\% mucin from the Hill lab, representative of a healthy patient. The sample was prepared according to the procedure presented in \cite{matsui2006physical}.  The ACF lacks a clear signal of dependency between increments.  Figure \ref{compare} will demonstrate that this data exhibits subdiffusivity, which cannot be detected here. }
	\end{center}
\end{figure}

For the purpose of capturing subdiffusivity in the data, it is worthwhile shifting the perspective from the time domain, typical of the currently employed heuristic methods (see (\ref{e:regression})), to the Fourier (frequency) domain. The latter is often superior for the characterization of the \textit{long term} memory  of stochastic systems.  Given sufficiently fine data, \emph{ i.e.} high resolution in space and time, it should be theoretically possible to extract this memory information from the increment process of the position data.  However, the fundamental limitations of the camera frame rate prevent the collection of adequately fine data for characterization using the autocovariance function.  For example, in Figure \ref{mucusacf} one sees virtually no signal in the sample autocorrelation of the increment process. 

The basis for spectral domain inference for long-term subdiffusivity is the connection between the spectral properties of the velocity process $V$ and the population quantity of primary interest, namely $G^*(\omega)$. More specifically, one can relate the large time MSD exponent $\alpha$ in (\ref{e:asympt_exp}) to the behavior of the spectral density of the velocity process at the origin, and in turn relate this to the behavior of the complex shear modulus $G^*(\omega)$ near the origin.

We assume that the spectral density $\widehat{\rho}(\omega)$ obeys a positive power law near the origin, i.e.,
\begin{equation}\label{e:behavior_f(x)_as_x->0}
\lim_{\omega \rightarrow 0} \frac{\widehat \rho(\omega)}{|\omega|^{2d}} = C > 0, \quad 0 < d < \frac{1}{2}.
\end{equation}
As is done in the long range dependence literature, we use the fractional parametrization $d$ out of notational convenience, and also because it naturally connects with the Hurst parameter when applicable (see expressions (\ref{e:d=H-1/2}) and (\ref{e:modified_g}) below). Under a mild assumption, one can show that
\begin{equation}\label{e:long_term_EX^2/t^alpha}
c  \leq \frac{\E{X^2(t)}}{t^{1-2d}} \leq  C, \textnormal{ for large $t$},
\end{equation}
for two constants $c,C > 0$ (see the Appendix). In particular, if there exists $0 < \alpha < 1$ such that (\ref{e:asympt_exp}) holds for large $t$, then
\begin{equation}\label{e:alpha=1-2d}
\alpha = 1 - 2d,
\end{equation}
which establishes the connection between the spectral density and subdiffusivity parameters. In particular, the velocity particle of a GLE driven by fBm is an example of a process whose spectral density satisfies (\ref{e:behavior_f(x)_as_x->0}) (see Section \ref{s:fGLE_integ_repres}).

We still need to explicitly connect the subdiffusive exponent and by extension, the parameter $d$ to the complex shear modulus $G^*$. Recalling that $\widehat \rho(\omega) \propto 1 / \eta^*(\omega) = i \omega / G^*(\omega)$ and putting this together with \eqref{e:alpha=1-2d}, we have the simple relationship between the asymptotic MSD exponent $\alpha$ and the observed behavior of the complex shear modulus $G^*$. Thus, when there is a clear asymptotic MSD exponent $0 < \alpha < 1$ we have
\begin{equation} \label{e:G-near-zero}
	\lim_{\omega \rightarrow 0} \frac{G^*(\omega)}{|\omega|^{\alpha}} = C' > 0.
\end{equation}
For perspective, we draw attention to Figure \ref{f:masonweitz}, which contains storage and loss moduli measurements from the original MW \cite{mason:weitz:1995} work. The subdiffusive exponent exactly corresponds to the behavior of the storage and loss moduli as the frequency approaches zero.


In reality, experimental data sets are made up of discrete measurements of the position process $X(t)$, not of the velocity $V(t)$. We should therefore consider the increment process $Y_j= \nabla X(\Delta j) = X(\Delta j) - X(\Delta (j-1) )$ associated with the position process $X(t)$. To simplify the following expressions, we will assume that $\Delta$ is one; this is equivalent to reparameterizing the problem into time units of the sampling increment. It turns out that, under assumption (\ref{e:behavior_f(x)_as_x->0}) and an additional mild condition, $Y_j$ inherits the fractional behavior of $V(t)$. In other words, let $\widehat{\rho}_{Y}$ be the spectral density of the discrete increment process $Y$. Then
\begin{equation}\label{e:behavior_fY(x)_as_x->0}
\lim_{\omega \rightarrow 0}\frac{\widehat{\rho}_{Y}(\omega)}{|\omega|^{2d}} = C > 0
\end{equation}
for some constant $C$; see the Appendix. The property (\ref{e:behavior_fY(x)_as_x->0}) implies that the discrete increment process $Y$ is \textit{antipersistent} and that it tends to display negative correlation across short time lags.

The expressions (\ref{e:alpha=1-2d}) and (\ref{e:behavior_fY(x)_as_x->0}) imply that a natural first approach to the problem of characterizing long term subdiffusivity is via spectral \textit{semiparametric }methods. Indeed, the estimation of the parameter $d$ does not necessarily involve the estimation of the entire function $\widehat \rho(\omega)$, or the specification of a full parametric model for $V(t)$. Instead, one only needs to look at the behavior of $\widehat \rho(\omega)$ close to the origin, a common idea from the literature on long memory processes. The fractional property (\ref{e:behavior_fY(x)_as_x->0}) opens the possibility of Fourier or wavelet-domain methods such as the Local Whittle estimator (K\"{u}nsch \cite{kunsch:1987}, Robinson \cite{robinson:1995}), spectral regression (Robinson \cite{robinson:1995:specreg}) or wavelet regression in the fashion of Veitch and Abry \cite{veitch:abry:1999}, Moulines et al.\ \cite{moulines:roueff:taqqu:2007}, Bardet \cite{bardet:2000,bardet:2002} (see also Beran \cite{beran:1994} and Palma \cite{palma:2007} for surveys). In particular, the Local Whittle is a well-studied semi-parametric estimator defined as
\begin{equation}\label{e:LocWhit_def}
\textnormal{argmin}_{d \in \Theta} \log\left( \frac{1}{m} \sum^{m}_{j=1} \frac{I_n(\omega_j)}{\omega^{2d}_j}\right) + 2d \frac{1}{m}\sum^{m}_{j=1}\log(\omega_j), \quad m \ll n,
\end{equation}
where $\Theta = [d_1,d_2]$, $0 < d_1 < d_2 < 1$. In (\ref{e:LocWhit_def}), $I_n(\cdot)$ is the periodogram, obtained from the discrete Fourier transform of a discrete sample path $Y_1, Y_2, ..., Y_n $ as
$$
I_n(\omega_j) := \frac{1}{2 \pi n} \Big| \sum^{n}_{k=1}Y_k e^{-i k \omega_j}\Big|^2, \quad \omega_j = \frac{2 \pi j}{n}, \quad j = 1,..., \Big[\frac{n-1}{2}\Big].
$$
The Local Whittle makes mild assumptions on the underlying process, i.e., that it is linear with innovations that are martingale differences satisfying a uniform integrability condition (Robinson \cite{robinson:1995}, pp.\ 1633-1634). There is the additional tuning parameter $m$, which indicates the number of frequencies used in the estimation. This tuning parameter is necessary because the spectral density of the process is semi-parametrically assumed to follow a power law (\ref{e:behavior_f(x)_as_x->0}) close to zero. So, the use of frequencies which are too distant from the origin could then introduce undesired information from other parts of the spectrum of the process. The Local Whittle estimator is asymptotically normal, and combined with expression (\ref{e:alpha=1-2d}) this implies that
\begin{equation}\label{e:LocWhit_alpha_asympt}
\sqrt{m}(\overline{\alpha} - \alpha) \sim N(0,1),
\end{equation}
where $m$ satisfies $1/m + m/n \rightarrow \infty$ as the sample size $n$ goes to infinity, which allows for asymptotic hypothesis testing.

\begin{figure}
	\begin{center}
	\includegraphics[height=2.5in,width=3in]{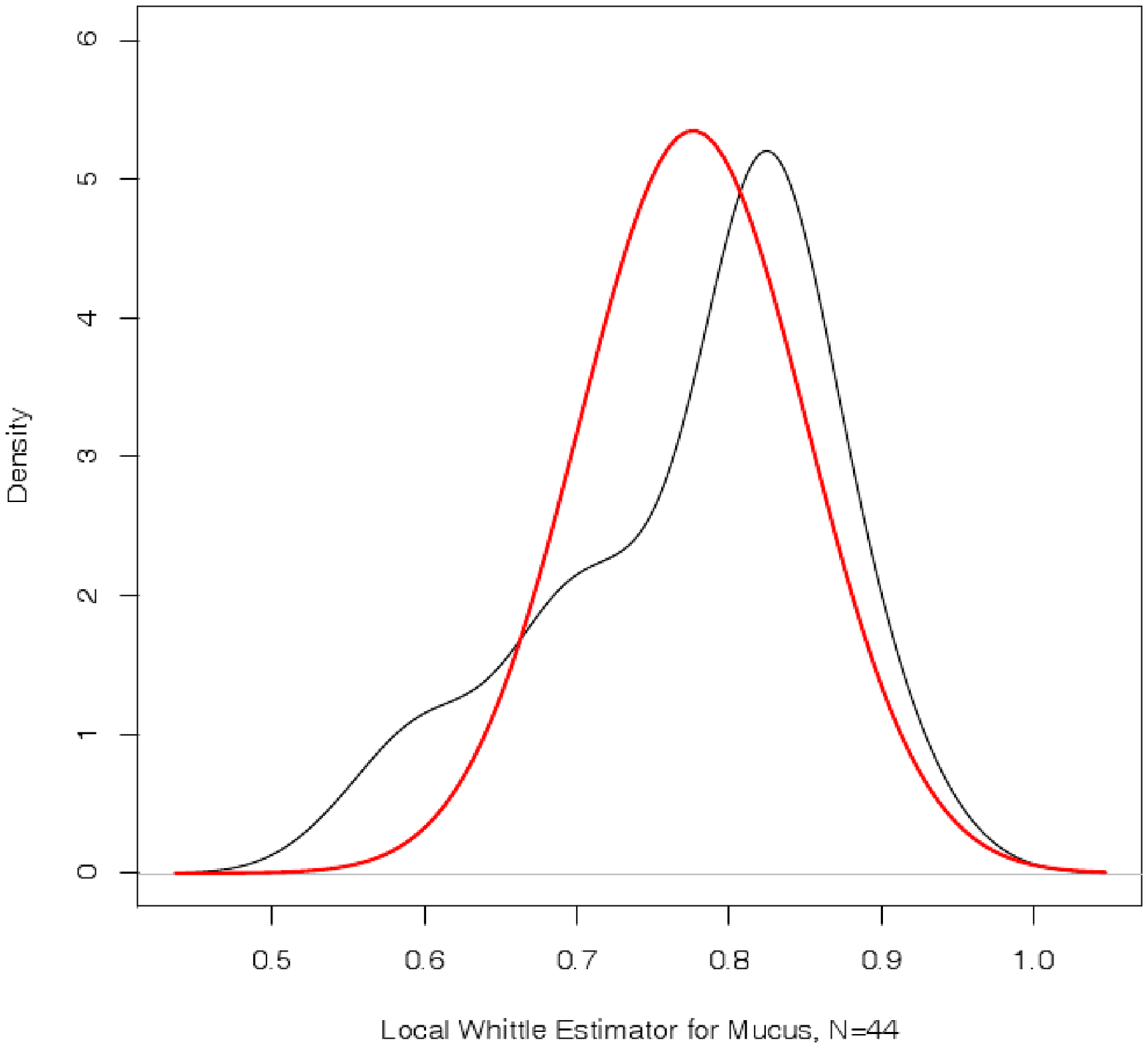} \ \includegraphics[height=2.5in,width=3in]{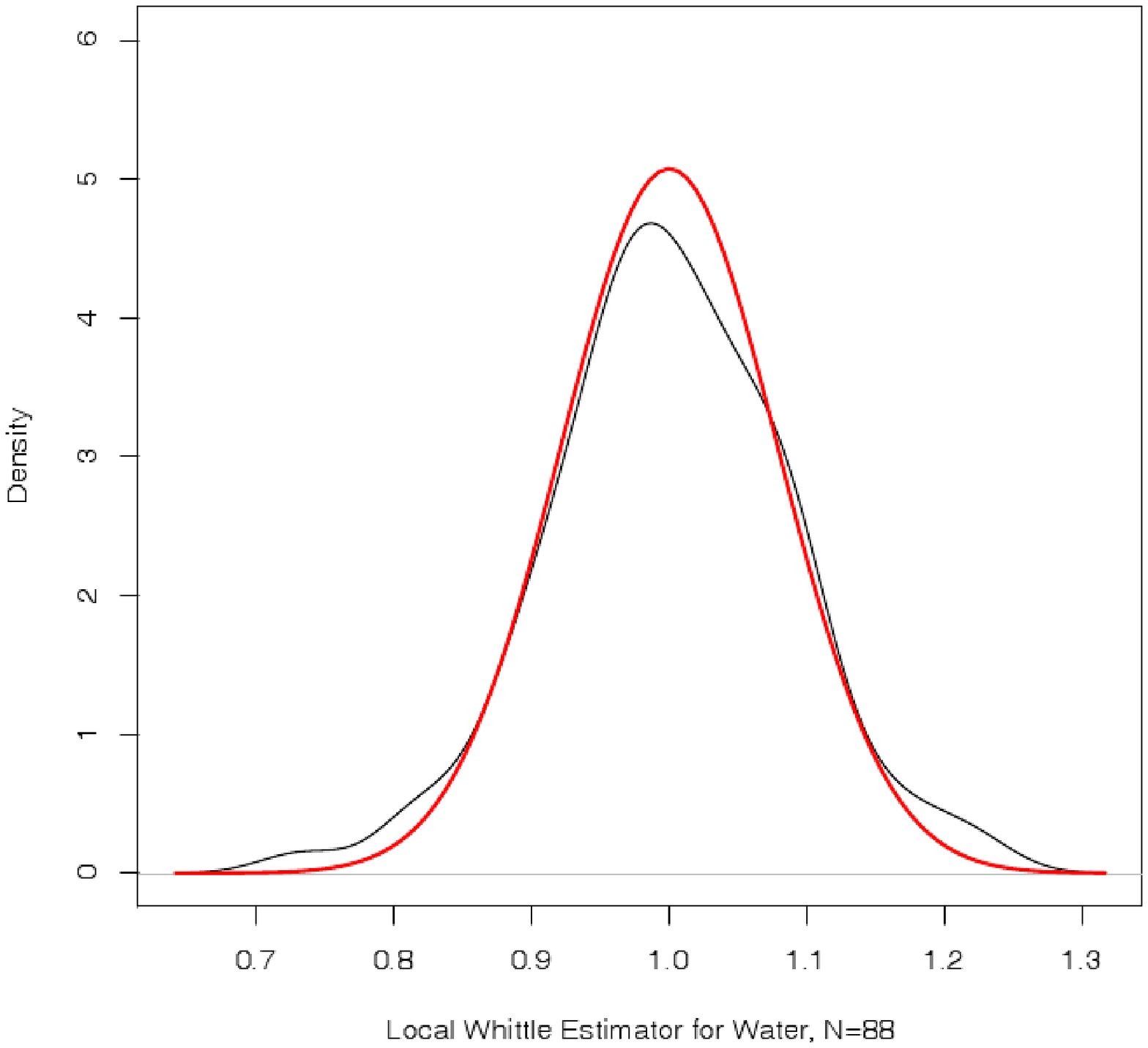}
	\caption{\label{compare} Estimated distribution of the Local Whittle estimator over various sample paths (data from the David Hill laboratory, UNC). The red line is a Gaussian curve centered at the local Whittle estimate from the pooled data.  The black line is a kernel density estimate summarizing the local Whittle estimates across multiple paths. Left plot: mucus. Right plot: water.}
	\end{center}
\end{figure}

\begin{figure}
	\begin{center}
	\includegraphics[height=2.5in,width=3in]{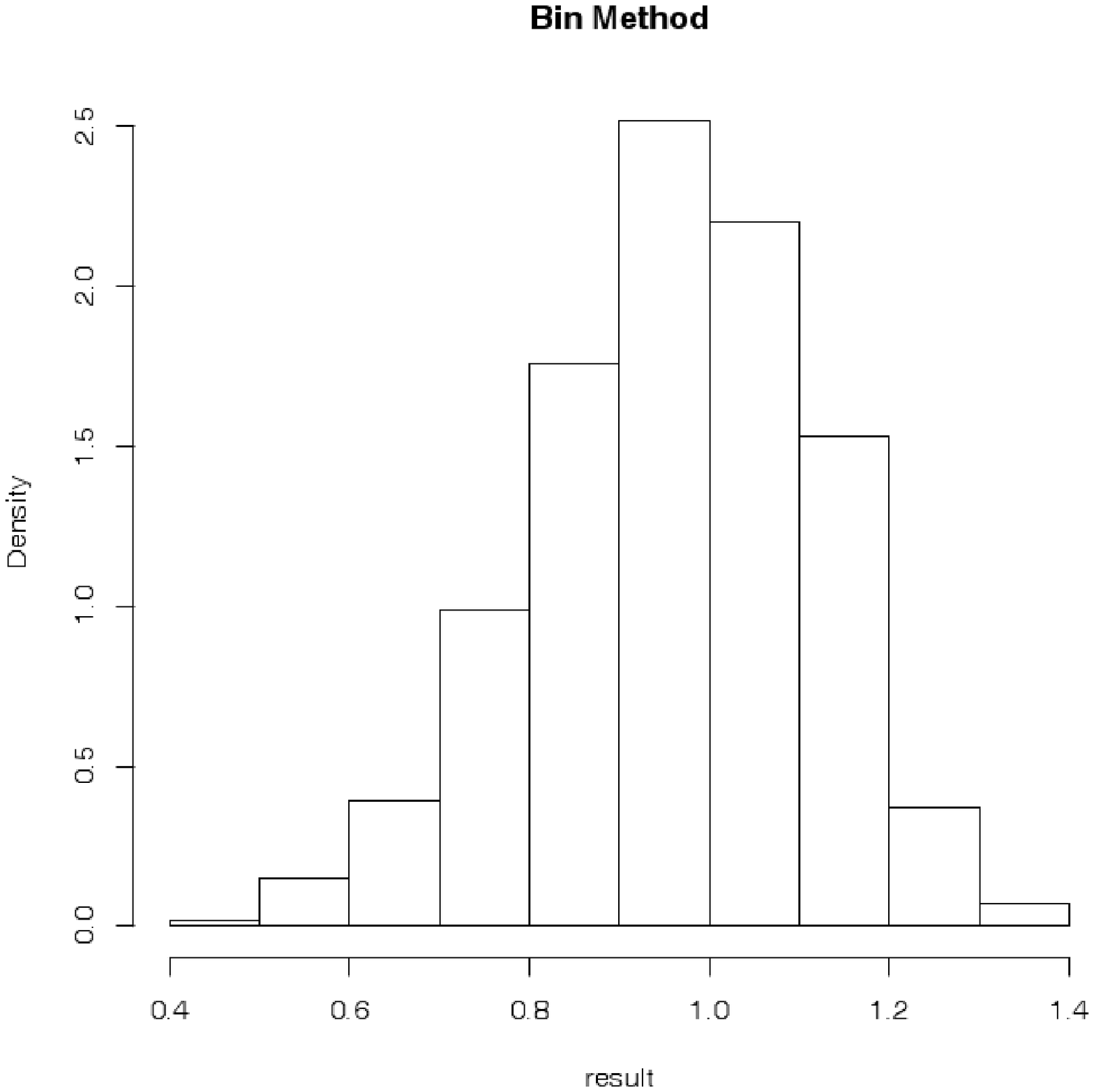} \ \includegraphics[height=2.5in,width=3in]{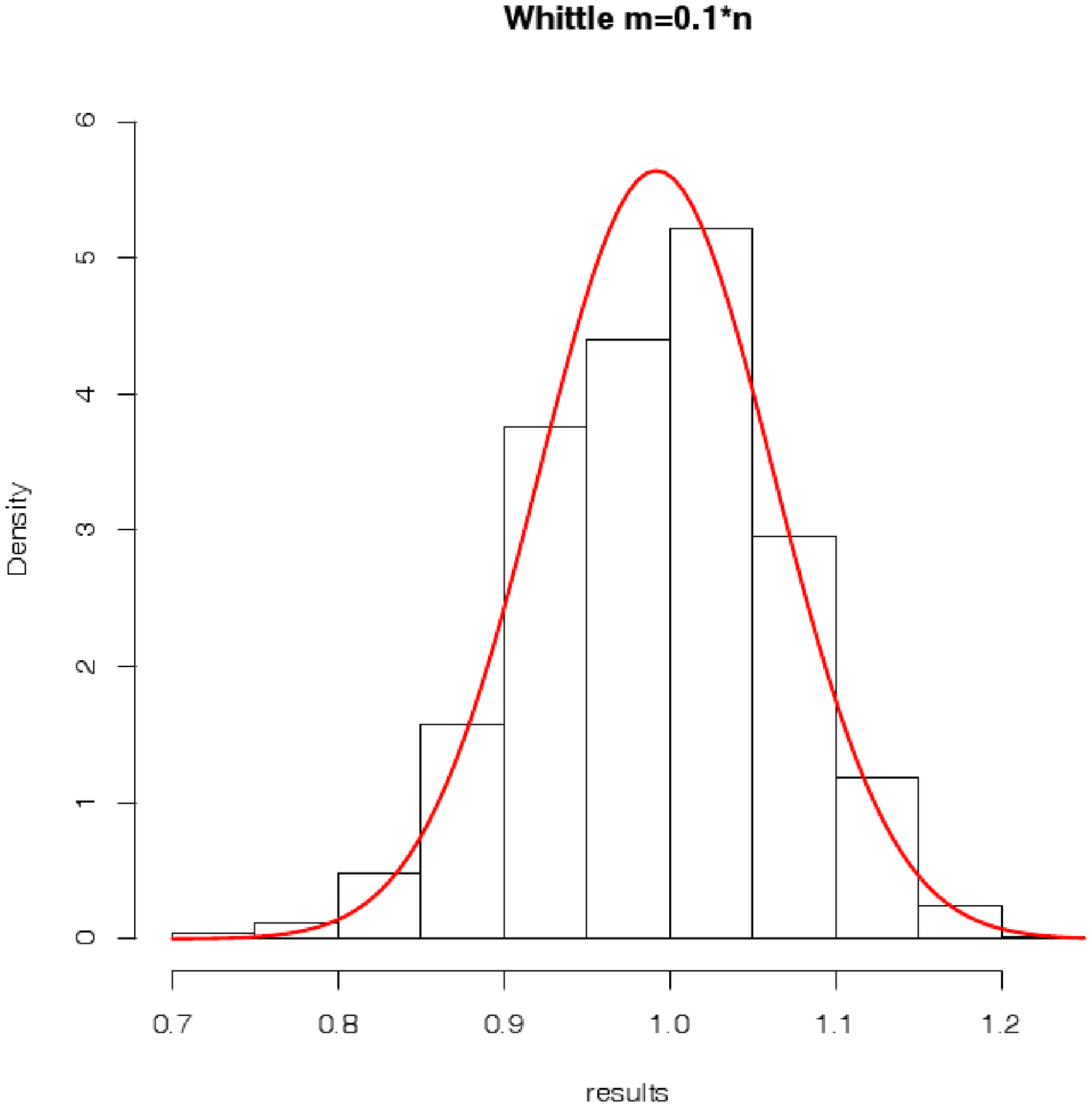}
	\caption{\label{Brownian_alpha_hat} Estimation of the subdiffusivity parameter $\alpha = 1$ for Brownian motion (path length 5,000; only the first 1000 lags are used). Left plot: pathwise MSD. Right plot: Local Whittle.   The red line in the Local Whittle plot represents the theoretical sampling distribution centered at the pooled estimate for $\alpha$. }
	\end{center}
\end{figure}

The right plot in Figure \ref{Brownian_alpha_hat} displays the simulated distributions of the pathwise MSD and the Local Whittle estimator for Brownian motion. Both are centered at about the right value $\alpha = 1$, but the latter exhibits a substantially smaller dispersion. Moreover, preliminary results with experimentally collected data indicate a substantial discrepancy in the distribution of the Local Whittle estimator between the cases of mucus and water (see Figure~\ref{compare}). This stands in stark contrast with the results obtained from the sample autocorrelation function, displayed in Figure \ref{mucusacf}.

\subsection{Challenges in Statistical Inference.}

There are a number of outstanding challenges in statistical inference for microrheology.  One important factor in the analysis and modeling of microrheological data is the relatively complex nature of the experiments. In general, the tracer beads cannot be observed precisely. The primary data resulting from a microrheological experiment is recorded in the form of a digital video. The latter is preprocessed using off-the-shelf software such as Spot Tracker and the NIH-funded ImageJ to identify and track the location of each bead in the complex fluid sample. Consequently, separating out the errors generated by image analysis is a considerable statistical task in its own right. 

Another important challenge is drift removal.  In a microrheological experiment, a drift can be induced by multiple factors including a possible heat gradient from the microscope lighting.  It has yet to be determined whether the drifts felt by particles are different, the same, or possibly random.  When developing methods to estimate either the memory kernel or mean squared displacement, one should at least consider the effects of such drift.  By using standard exploratory tools, the authors were able to identify numerous anomalies in the data emerging from David Hill's laboratory such as non-linear drifts and mis-tracked beads which allowed the laboratory to refine their experimental data collection.  It has been an ongoing challenge to find data sets which are sufficiently unaffected by such experimental issues to properly apply some of the theoretical tools that are being developed.


There is also the issue of the detection of heterogeneity, or non-stationarity, in the (sub)diffusive behavior of beads within experiments.  Biofluids such as mucus and cytoplasm have a highly complex chemical composition with at least 250 different chemical component which may contribute to such heterogeneity. Do different beads behave differently within the same physical sample? Is there spatial variation in the diffusive behavior for the same bead? The former could indicate variations in the way beads interact with the material, for example due to surface chemistry; the latter may indicate heterogeneity of the material under study, which may be difficult to discern and quantify visually.  An outstanding issue is the modeling and inference for large sets of tracer beads taken within the same experiment (i.e., physical sample or locus), or across different paths.  One could consider adapting  ``longitudinal" methods from the time series literature to quantify heterogeneities across paths, such as those found  in Shumway and Stoffer \cite{shumway:stoffer:2006} and Durbin and Koopman \cite{durbin2001time} in combination with the Kalman approach of Fricks et al~\cite{fricks:yao:elston:forest:2009}.

One important question is whether the GLE is the right model or even the right type of model.  Some data has suggested that particles bind to the polymer network and then break free to then bind again after diffusing some distance~\cite{saxton1996anomalous}.  These type of models go by several names depending on the field: regime switching in economics, Markov modulated processes in applied probability, and a flashing ratchet in some sub-disciplines of physics.   There are a number of methods to handle this type of switching behavior statistically~\cite{shumway:stoffer:2006,boysen2009consistencies,kim1999state}.  However, one possibility is that particles exhibit subdiffusion when free from the polymer network, and this would present new challenges.

Another challenge is to test and perhaps drop the implicit assumption in the literature that the two spatial dimensions of the bead's location $(X_1(t),X_2(t))$ are independent. One natural way to do this is to develop a new GLE-like model driven by the multivariate process operator fBm (e.g., Didier and Pipiras \cite{didier:pipiras:2011}) and test it based on recently-developed multivariate estimators (e.g., Abry and Didier \cite{abry:didier:2011}).


\section{Stochastic Simulation.} \label{simulation}


\subsection{Methods of Simulation.}

%
%

There are various assumptions under which the GLE yields stationary solutions, either for the position process $X(t)$ or the velocity process $V(t)$.  Stationarity allows for the application of a wide range of techniques designed for such processes under the assumption of Gaussianity; this is especially convenient since the system is in general non-Markovian.

As described in \cite{bardet:lang:oppenheim:philippe:taqqu:2003}, there are various ways to simulate a Gaussian process: ($S1$) if the covariance is known, by means of a sequence of variables whose covariance is the given one; ($S2$) based on a representation of the process (such as a time or spectral domain stochastic integral), possibly after some simplification (e.g., truncation) of this representation; ($S3$) through a discrete sequence whose weak limit is the target process. 

With regard to ($S3$), if a natural discretization scheme is available, simulating a continuous-time process is potentially straightforward. A GLE is a kind of SDE. In the case of It\^{o} diffusions driven by Brownian motion, Euler-type schemes, which rely on discrete approximations at equidistant grid points, can be used for simulation. For instance, see \cite{asmussen:glynn:2000} or \cite{kloeden1995stochastic}. In general, for SDEs which are driven by other types of Gaussian noise, the theory is not as well-developed; on the pathwise approximation to SDEs driven by fBm, for instance, see \cite{neuenkirch:2008} and references therein.  However, it is well-known that an equally-spaced sampling of the OU process produces an AR(1) sequence. Thus, a fast and exact simulation procedure is to generate an AR(1) vector with appropriate parameter values. In \cite{fricks:yao:elston:forest:2009}, the multivariate equivalent of this fact is used to simulate the solution to a generalized Langevin equation when the memory kernel is a Prony series.  A related example can be found in \cite{mckinley:yao:forest:2009}, where the authors show that the same Prony series GLE can be represented as a Brownian motion plus the sum of OU processes when taking a zero mass limit.  In general, though, the correlation structure of discretized continuous time processes, such as the GLE, is complex and not amenable to a simulation scheme based on a simple loop, as with the OU process.

The availability for the GLE of integral representations or covariance functions of the form (\ref{eq:gle-v-soln}) and (\ref{eq:rho-fourier}), respectively, indicates that it is natural to pursue simulation methods that fall under ($S1$) or ($S2$). We now describe two such simulation methods which assume known covariance functions and that will be of interest in the simulation of GLEs. Later on, in Subsection \ref{s:wave_sim}, we will also describe a wavelet-based method that falls under ($S3$), but which builds upon integral representations.

So, let $X$ be a target zero mean Gaussian stationary process with a given, known covariance matrix $\Sigma$ over a finite set of time points $N$.  One classical and simple way to simulate $X$ consists of performing a Cholesky decomposition of $\Sigma = LL^{*} $, where $L$ is a lower triangular matrix, and generate a vector $Z$ of i.i.d.\ standard Gaussian variables. Then $X \stackrel{d}= LZ$, as desired. Cholesky-based simulation is exact up to computational error and can be implemented recursively (see \cite{asmussen:glynn:2000}, pp.\ 311-312). However, it is slow in terms of computational complexity: $O(N^3)$, where $N$ denote the length of the resulting stochastic vector.

Another popular method is the Circulant Matrix Embedding (CME; see \cite{davies:harte:1987}, \cite{wood:chan:1994}, \cite{dietrich:newsam:1997}; see \cite{johnson:1994}, \cite{beran:1994} and \cite{asmussen:glynn:2000} for reviews of this and other methods).  The algorithm draws upon embedding the covariance matrix in a non-negative definite circulant matrix of size $M \geq 2(N-1)$. This is computationally convenient, since the diagonalization of circulant matrices can be carried out by means of the Fast Fourier Transform (FFT), which has complexity $O(N \log(N))$. Like Cholesky-based simulation, CME is exact. For the reader's convenience, we summarize the CME. Let ${\mathcal F}$ be the FFT. The simulation procedure is as follows:
\begin{enumerate}
\item [$(C_1)$:] choose $M = 2^p \geq 2(N-1)$;
\item [$(C_2)$:] form the vector $\Lambda := (\lambda(0),\lambda(1),...,\lambda(M/2-1),\lambda(M/2),\lambda(M/2-1),..., \lambda(1))$, where $\lambda(h)$ is covariance function evaluated at the time lag $h$;
\item [$(C_3)$:] compute $\widehat{\Lambda}:={\mathcal F}(\Lambda)$. Check whether the entries of $\widehat{\Lambda}$ are all non-negative. If so, proceed to the next step. If not, choose a larger $p$;
\item [$(C_4)$:] generate $Z := X+iY$ where $X,Y \stackrel{\textnormal{i.i.d.}}\sim N(0,I(1/\sqrt{2}))$;
\item [$(C_5)$:] define $U$, $U_k := \sqrt{\widehat{\Lambda}_k}Z_k$, $k = 1,..., M$;
\item [$(C_6)$:] compute $X := \Re( {\mathcal F}^{-1}(U) )$.
\end{enumerate}
The vector $X$ obtained is Gaussian and has the desired covariance structure. Of course, the non-negativity condition in $(C_3)$ is essential. Although no general theoretical results are available, it has been justified for many models of interest. For instance, it is shown in \cite{dietrich:newsam:1997} that a sufficient condition is that the sequence of covariances $\lambda(0), \lambda(1), ..., \lambda(N-1)$ be non-negative, decreasing and convex (see also \cite{asmussen:glynn:2000}, \cite{percival:constantine:2002}, \cite{craigmile:2003}). One limitation of the CME is that it is not iterative in the sense that it requires the simulation horizon to be predefined.


\subsection{The fractional GLE: correlation structures via integral representations}\label{s:fGLE_integ_repres}

In order to provide a study of simulation of GLE-based anomalous diffusion, we focus on the fBm-driven GLE (fGLE), since its correlation structure has been well-studied. In this section, based on \cite{kou:2008}, we take a closer look at the correlation structure of the fGLE via integral representations (see also Section \ref{sec:GLE}).

\begin{figure}
	\begin{center}
	\includegraphics[height=2.5in,width=3in]{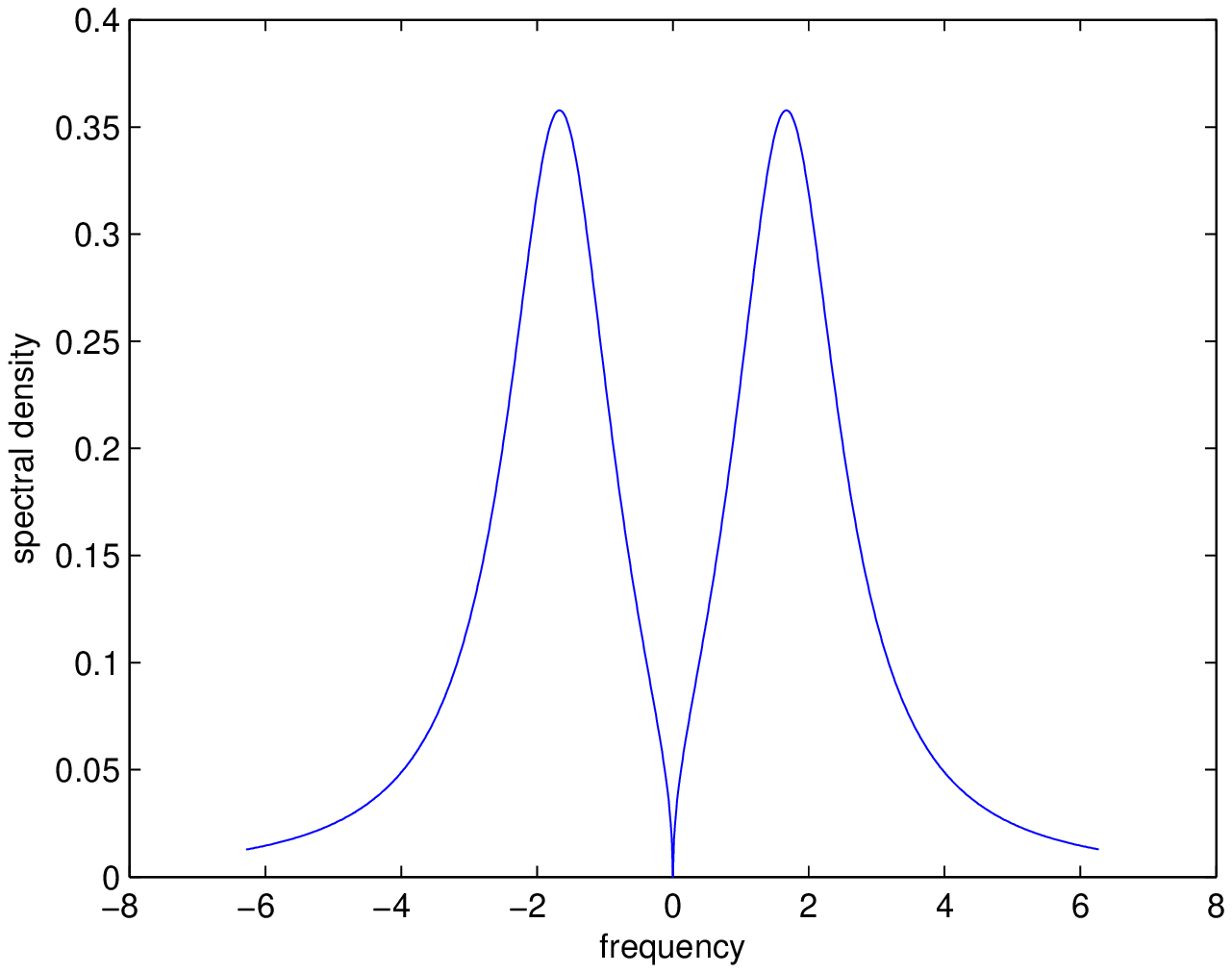} \ \includegraphics[height=2.5in,width=3in]{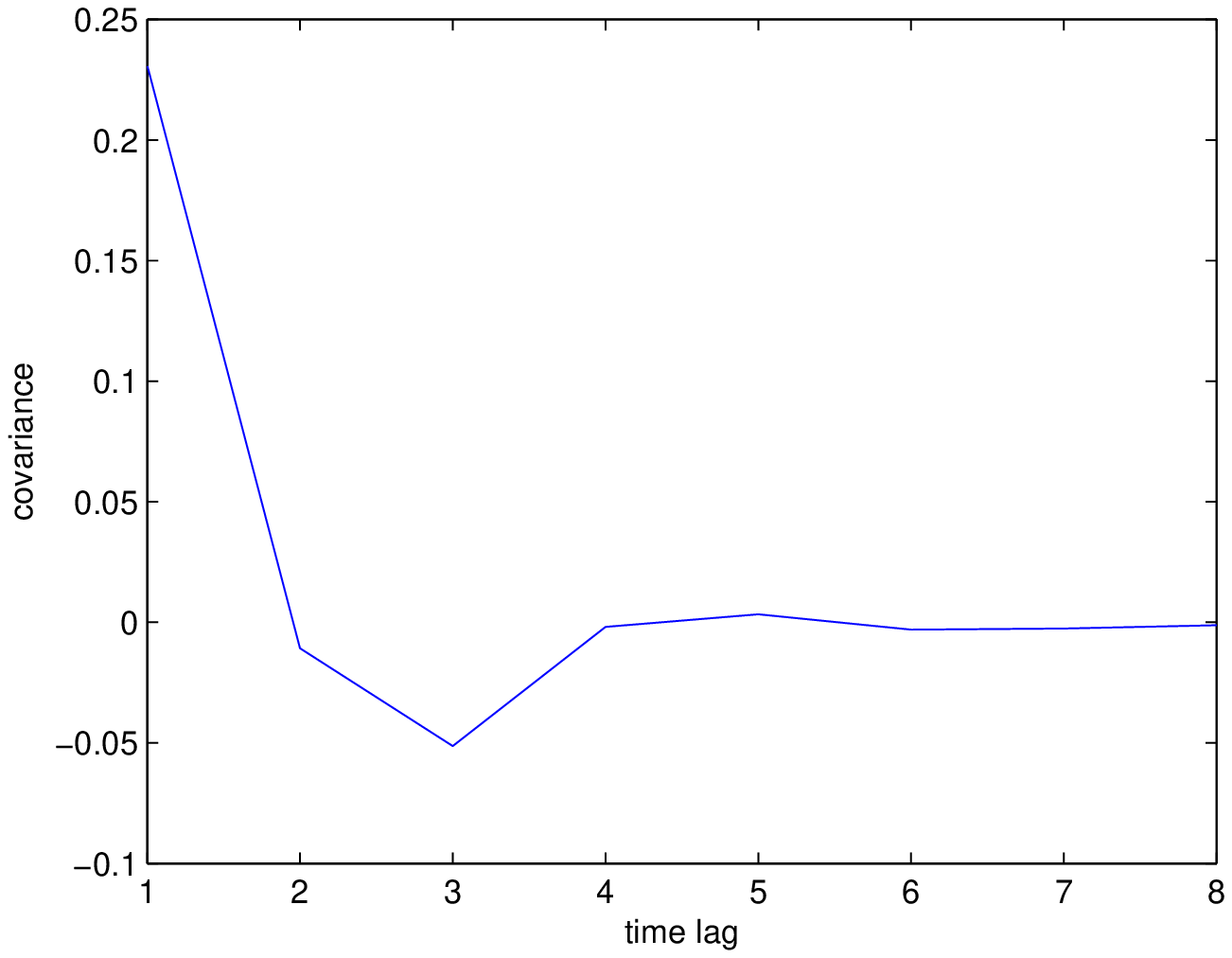}
	\caption{\label{f:fracGLE_correl_structure} fGLE, correlation structure up to a multiplicative constant ($d =0.25$, $\gamma = 2$, $ m =1$). Left plot: spectral density. Right plot: covariance function.
	}
	\end{center}
\end{figure}

In that paper, the author analyzed the fGLE in three situations: when it models a free particle, when the particle is subject to the harmonic potential $U(x)= \frac{1}{2}m \psi x^2$, and when the acceleration term $m dV(t)/dt$ is negligible, corresponding to the overdamped condition in physics, where the friction is quite large. Because it is the most relevant for passive microrheology, we will focus on the free particle system, i.e.,
\begin{equation}\label{e:fGLE}
m dV(t) = - \gamma \Big( \int^{t}_{-\infty}V(u) \Gamma_H(t-u)du\Big) dt + \sqrt{2\gamma k_B T}dB_{H}(t).
\end{equation}
Kou \cite{kou:2008} gives the solution $V(\cdot)$ to \eqref{e:fGLE} as a pathwise defined Riemann-Stieltjes integral. In the spectral domain, one can develop such solution into the integral representation
\begin{equation}\label{e:V_spec_integ_repres}
V(t) \stackrel{\mathcal L}= \frac{\sqrt{2 \gamma k_B
T}}{c(d)}\hspace{1mm}\int_{\bbR} e^{it \omega}\frac{1}{\gamma
\kappa(\omega)|\omega|^{-2d} - im\omega} |\omega|^{-d} d\widetilde{B}(\omega), \quad 0 < d
< \frac{1}{2},
\end{equation}
where
\begin{equation}\label{e:d=H-1/2}
d := H - 1/2,
\end{equation}
$\gamma$ is the friction constant, $k_B$ is the Boltzmann constant, $T$ is the underlying temperature, $c(d)$ is a constant, $\widetilde{B}(\omega):=\widetilde{B}_1(\omega)+i\widetilde{B}_2(\omega)$ for the real-valued Brownian motions $\widetilde{B}_1$, $\widetilde{B}_2$ satisfying $d\widetilde{B}(-\omega) = \overline{d \widetilde{B}(\omega)}$ a.s. and
$$
\kappa(\omega) = \Gamma(2H + 1)(\sin(H \pi) - i \textnormal{sign}(\omega) \cos(H \pi) ).
$$
We can rewrite the spectral filter of $V$ with respect to the complex Brownian measure $d\widetilde{B}(\cdot)$ as
\begin{equation}\label{e:modified_g}
\widehat{g}(\omega) = \frac{1}{(\nu_0 + \nu_1 |\omega|^{\beta} + \nu_2 |\omega|^{2 \beta})^{1/2}} |\omega|^{d}, \quad \beta := 1 + 2d,
\end{equation}
for appropriate constants $\nu_0$, $\nu_1$, $\nu_2$. Expression (\ref{e:modified_g}) is the basis for the simulation of fGLE in Subsections \ref{s:wave_sim} and \ref{s:eval_sims}.
 In particular, expression (\ref{e:V_spec_integ_repres}) implies that the fGLE is antipersistent (see (\ref{e:behavior_f(x)_as_x->0})). Figure \ref{f:fracGLE_correl_structure} depicts its correlation structure in the spectral and time domains.

\subsection{Wavelet-based simulation}\label{s:wave_sim}


In this section, we describe a new, approximate wavelet-based simulation method for the fGLE. The discussion is based on \cite{didier:fricks:2011} and \cite{didier:pipiras:2008}, where the mathematical details are developed.

The simulation method goes under ($S3$), since it generates sequences of converging discrete time processes. This method builds upon approximate discretizations generated according to analytical and computational convenience. Moreover, it has the following features: it is computationally fast, potentially reaching complexity $O(N)$, since it is based on a Fast Wavelet Transform-like algorithm; it provides approximations that converge uniformly over compact intervals almost surely; the convergence speed is exponentially fast and depends on the sample path smoothness of the limiting process. Moreover, it is iterative both intensively and extensively. In other words, a generated approximation at scale $J$ over a compact interval $[0,T]$, $T \in \bbN$, can be used to generate a finer approximation at scale $J+1$ over $[0,T]$ or some expanded interval $[0,T + \tau]$, $\tau \in \bbN$.

The simulation method involves an appropriate sequence of filters $g_{j}$, $j=0,1,...,J$, where $J$ is the finest approximation scale chosen, and an associated sequence of low- and high-pass wavelet filters $u_j$ and $v_j$, respectively. The method then amounts to recursively generating an induced sequence of discrete approximations $V_j$, at scale $j \in \bbN$, of the continuous time velocity process $V$, where
$V_{j,k} = \sum^{\infty}_{n=-\infty} g_{j,n} \xi_{k-n}$, and $\{\xi_{k}\} \stackrel{\textnormal{i.i.d.}}\sim \textnormal{WN}(0,1)$.
The choice of the sequence of discretization filters $g_j$ is the key requirement. Heuristically, it should be such that
\begin{equation}\label{e:GJ_approx_1}
G_{j}(2^{-j}\omega) := \frac{\widehat{g}_{j}(2^{-j}\omega)}{\widehat{g}(\omega)}\approx 1, \quad j \rightarrow \infty, \quad \omega \in \bbR,
\end{equation}
where $\widehat{g}_{j}(\omega)$ is periodically extended to $\bbR$. Intuitively, the discretization filter approximates, up to a scaling factor,  the continuous time filter as the scale becomes finer and finer.

So, let $\uparrow_2$ be the upsampling by factor 2 operator, i.e., it inserts a zero between any two entries of a vector. The wavelet filters $u_j$ and $v_j$ are defined in the Fourier domain as
\begin{equation}\label{e:uj,vj}
\widehat{u}_j(\omega) := \frac{\widehat{g}_{j+1}(\omega)}{\widehat{g}_{j}(2\omega)}\widehat{u}(\omega), \quad \widehat{v}_j(\omega) := \widehat{g}_{j+1}(\omega)\widehat{v}(\omega),
\end{equation}
where $u$, $v$ are the conjugate mirror filters (CMF) of the underlying wavelet multiresolution analysis (see \cite{mallat:1999}, chapter 7). The simulation procedure can be briefly described as the following Fast Wavelet Transform-like algorithm.
\begin{enumerate}
\item [$(W_1)$:] at scale/step 0, generate via an exact method (e.g., CME) one first approximation sequence $V_0$;
\item [$(W_2)$:] at scale/step $j \in \bbN$, and given an approximation sequence $V_j$, obtain the next discrete approximation $V_{j+1}$ at scale/step $j+1$ via the relation
$$
V_{j+1} = u_j \ast \uparrow_2 V_j + v_j \ast \uparrow_2 \varepsilon_j.
$$

\end{enumerate}



The algorithm works because at step $j+1$, the FWT annihilates the correlation structure (kernel) $\widehat{g}_{j}$ of the approximation $\{V_{j,k}\}$ at scale $j$, and replaces it with a new, pre-chosen correlation structure $\widehat{g}_{j+1}$. This is a consequence of \eqref{e:uj,vj} and of the properties of the CMFs $u$ and $v$ (see also \cite{pipiras:2005}).

As a simulation procedure,  the resulting wavelet coefficients converge appropriately to the stochastic signal, i.e., $2^{J/2}V_{J,[2^{J}t]} \rightarrow V(t)$, where $[x]$ denotes the integer part of $x \in \bbR$. Such property is a consequence of (\ref{e:GJ_approx_1}) and of the properties of the scaling function $\phi$.
The choice of the function $G_{J}(\cdot)$ plays a key role. If $V$ is an OU process, then equally sampled points make up an AR(1) sequence. Therefore, for simulation purposes,
\begin{equation}\label{e:GJ_OU_exact}
G_{J}(\omega) = \frac{(1 - e^{-\lambda 2^{-J}}e^{- i \omega})^{-1}}{(\lambda + i \omega)^{-1}}
\end{equation}
is a natural choice, since $(1 - e^{-\lambda 2^{-J}}e^{- i \omega})^{-1}$ is a spectral filter of the associated AR(1) sequence. Moreover, the heuristic relation (\ref{e:GJ_approx_1}) is indeed satisfied.

However, exact and simple discretization schemes such as (\ref{e:GJ_OU_exact}) are not generally available. As a rule, the discretization of a continuous time process leads to a substantially more intricate expression for the spectral density. This is true, for instance, for fBm and the fGLE. In particular, for the latter, even though expressions for the covariance structure of the fGLE are available, they do not appear in closed form, which calls for numerical methods. 

So, a natural question is whether one could construct converging discretizations by means of a simple method that applies to a wider class of stochastic processes, in particular, the fGLE. Indeed, this can be done by developing non-causal discretization filters $\widehat{g}_{j}$ in three elementary steps:
\begin{itemize}
\item [($t.1$)] extend the truncated function $\widehat{g}(\omega)1_{[-\pi,\pi)}$ periodically to $\bbR$ (i.e., $\widehat{g}_j$ stems directly from $\widehat{g}$);
\item [($t.2$)] modify the resulting function with rescaling terms (e.g., $2^{-j}$) so that relation (\ref{e:GJ_approx_1}) holds;
\item [($t.3$)] smooth the resulting function at $-\pi$, $\pi$ so to speed up the time domain decay of the filter in theory and computational practice.
\end{itemize}

The method described in steps $(t.1) - (t.3)$ produces a sequence of discrete time processes $V_{j} = \widehat{g}_{j}\ast \varepsilon$ which is \textit{approximate} in two senses. First, because of numerical error including truncation, which is inevitable in computational practice with convolution-based procedures.  Second, because $g_j$ is picked for analytical and computational convenience, and thus $V_{j}$ is not in general an exact discretization of $V$. However, one can show that these approximate discretizations still converge exponentially fast to the correct limiting process.

\begin{figure}
	\begin{center}
	\includegraphics[height=2.5in,width=3in]{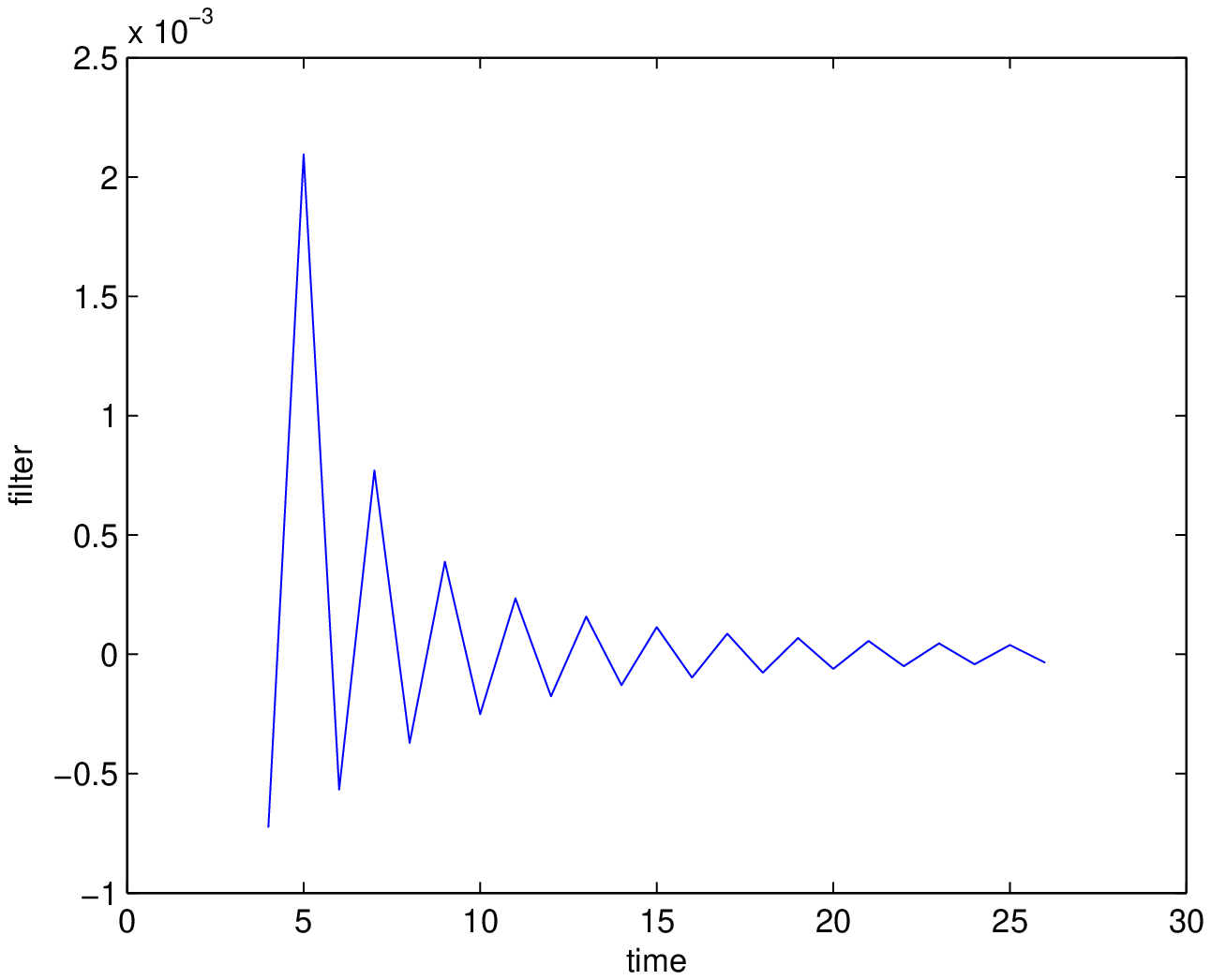} \ \includegraphics[height=2.5in,width=3in]{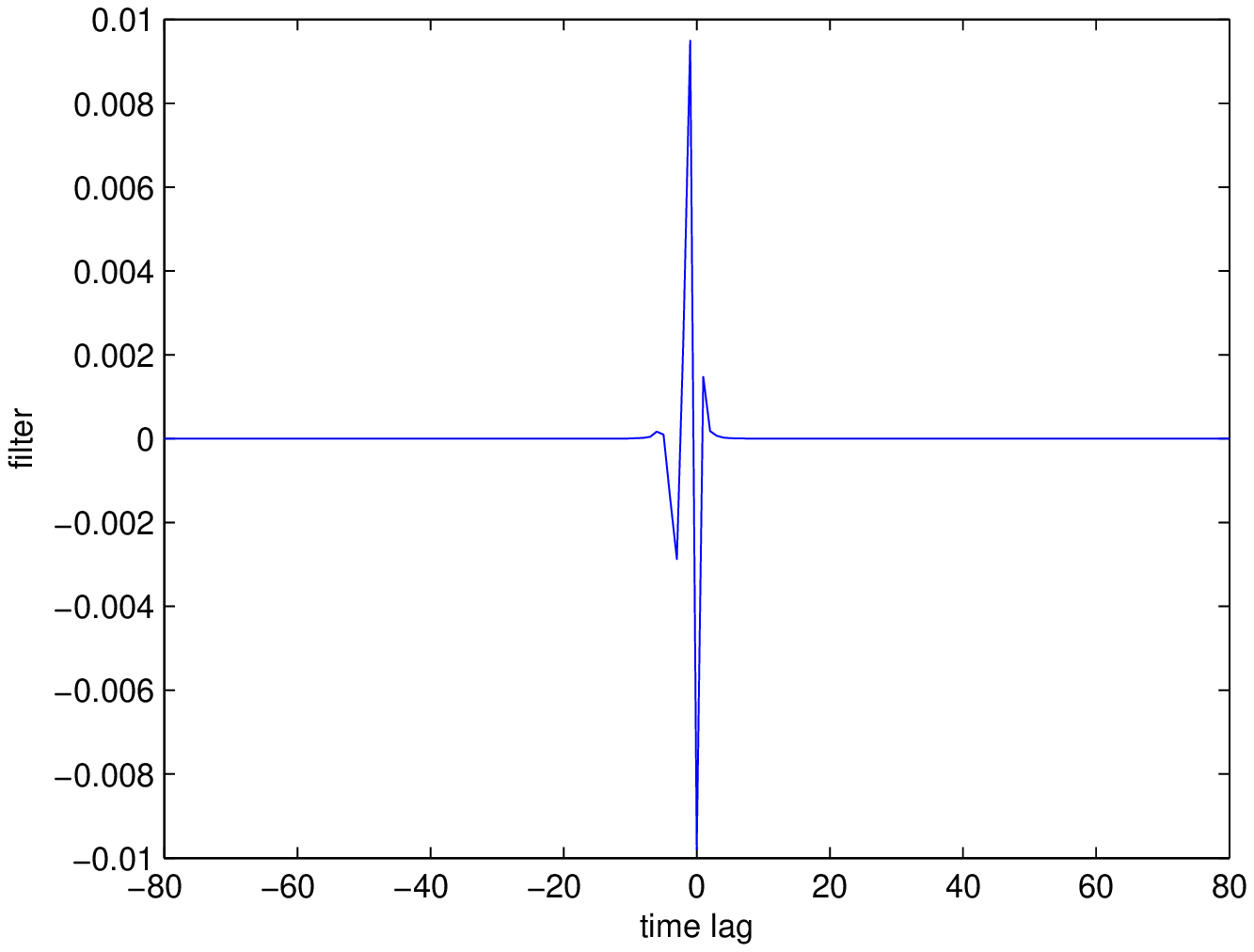}
	\caption{\label{f:fracGLE_v} Left: Non-smoothed high-pass filter, time domain. Right: fGLE, high-pass filter at $j=4$ ($\gamma = 2$, $ m =1$, $d =0.25$), 4 zero moments, time domain.
	}
	\end{center}
\end{figure}

By applying $(t.1)-(t.3)$ to \eqref{e:modified_g}, the proposed filter for the free-particle fGLE is
\begin{equation}\label{e:g_j_fGLE}
\widehat{g}_{j}(\omega) = \widehat{g}_{j,\nu,\beta}(\omega)\widehat{g}_{j,d}(\omega),
\end{equation}
where
\begin{equation}\label{e:fracOU_patched}
\widehat{g}_{j,d}(\omega) = 2^{jd} \exp\Big(\frac{-1}{2 \pi^2} \Big( \frac{\omega^{*}_d \omega}{\pi}\Big)^2 \Big) \Big| \frac{\omega^{*}_d \omega}{\pi}\Big|^{d}, \quad \omega^{*}_d:= \pi \sqrt{d}, \quad \omega \in [-\pi,\pi).
\end{equation}
\begin{equation}\label{e:g_j_gamma}
\widehat{g}_{j,\nu,\beta}(\omega) := \exp\Big(\frac{\beta}{2 \pi^2} \omega^2 \Big)  \frac{1}{(\nu_0 + \nu_1 2^{j \beta} |\omega|^{\beta}+ \nu_2 2^{2j \beta}|\omega|^{2\beta})^{1/2}}, \quad \omega \in [-\pi,\pi).
\end{equation}
 In \eqref{e:fracOU_patched} and \eqref{e:g_j_gamma}, the exponential terms smooth the original truncated filters. Figure \ref{f:fracGLE_v}, right plot, shows the designed high-pass filter in the time domain. The oscillations are almost invisible for relatively low lag values, and the actual decay is fast. Moreover, $\widehat{g}_{j}(2^{-j}\omega) \approx\widehat{g}(\omega)$ for large $j$, as expected. For the sake of comparison, Figure Figure \ref{f:fracGLE_v}, left plot, displays the time domain filter obtained without the smoothing terms.

%
%

\begin{figure}
	\begin{center}
	\includegraphics[height=2.5in,width=3in]{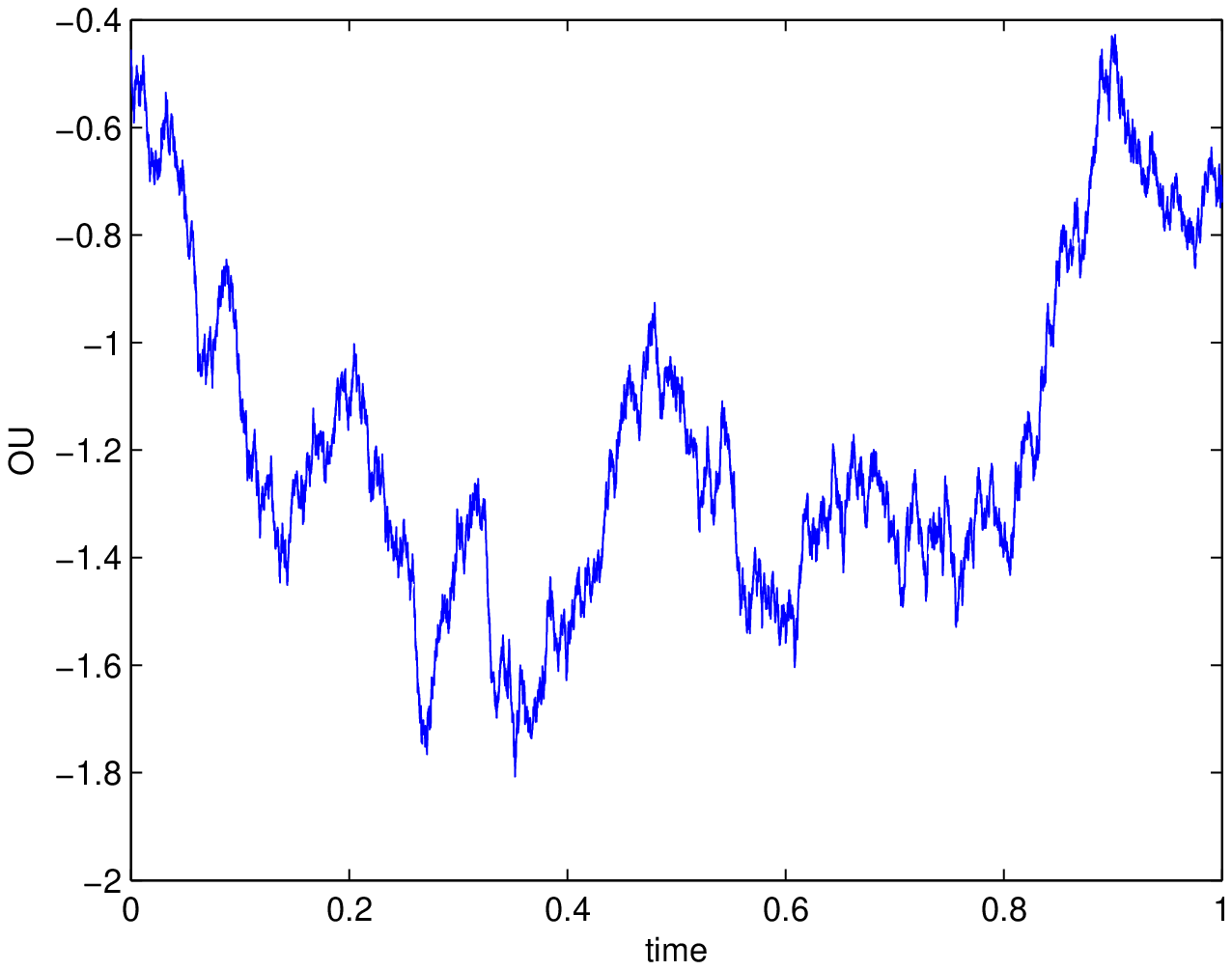} \ \includegraphics[height=2.5in,width=3in]{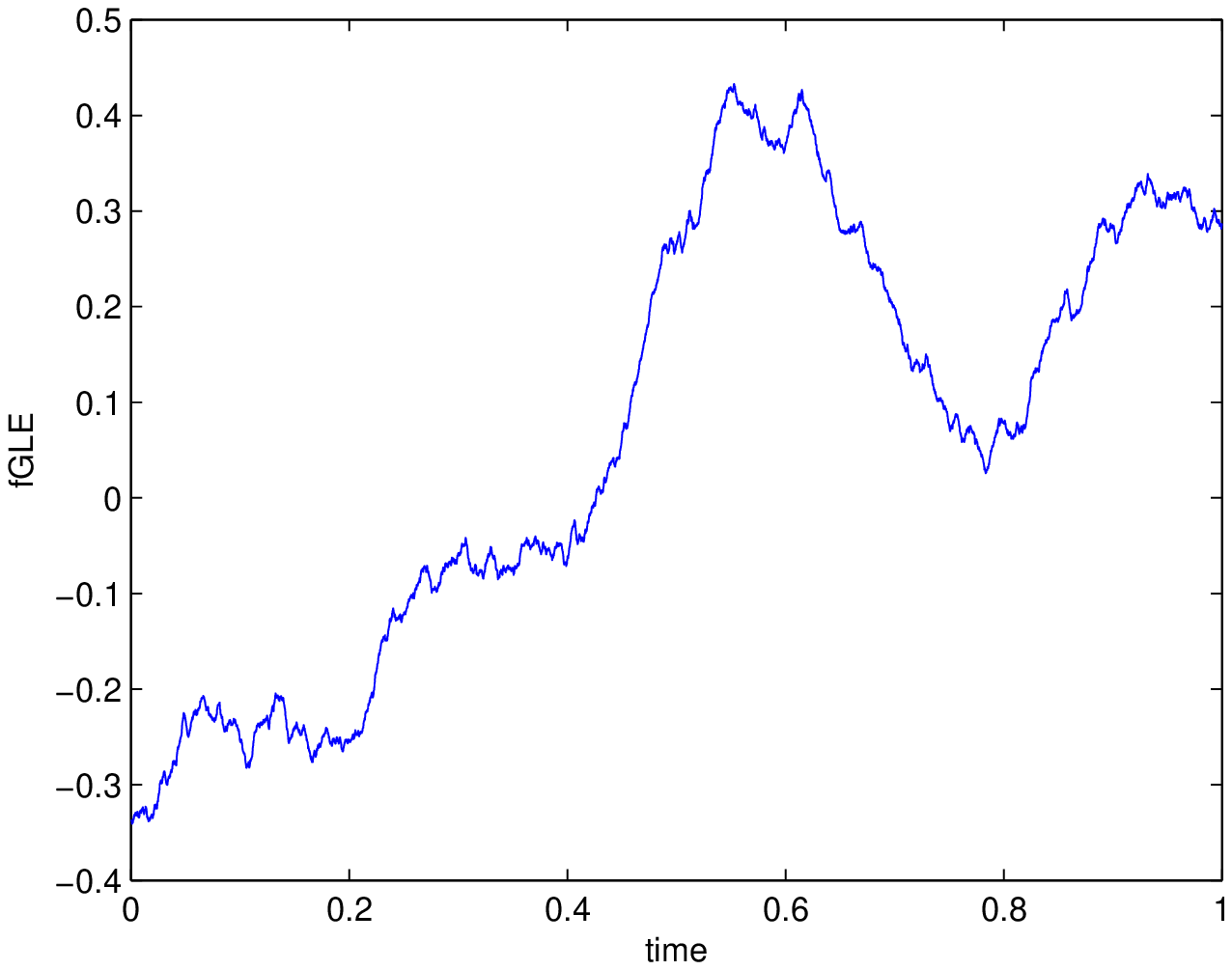}
	\caption{\label{f:OU_fracOU_samplepaths} Sample paths. Left: OU process, parameter value $a =1$ (4 zero moments). Right: fGLE process, parameter values $d =0.25$, $\gamma = 2$, $m =1$ (4 zero moments).
	}
	\end{center}
\end{figure}

\subsection{Evaluating simulation methods}\label{s:eval_sims}

Most simulation techniques are supported by theorems that establish some sort of convergence, equality in law, and so on. However, the finite sample performance, or accuracy, can in principle be disparate across methods. One idea is to use estimators in order to compare the simulation methods. Nevertheless, only the asymptotics of estimators are available in most cases, and the finite sample performance of estimators, e.g., bias, is studied based on simulation, which creates a circularity.

In view of this, we can look at measures of performance of the methods relative to one another. Since Cholesky-based simulation is a simple and exact procedure, we choose it to provide the baseline for the CME and the wavelet-based method (see also \cite{bardet:lang:oppenheim:philippe:taqqu:2003}, \cite{craigmile:2005}). A two-sample $t$ statistic is used to assess the difference between the values of the estimator when generated by two of the methods. We evaluate the quality of the simulation based on the Local Whittle estimation of the parameter $d$.
This is of special interest in our framework of subdiffusion, since the Local Whittle is a good estimator for the subdiffusivity parameter, as discussed in Section \ref{inference}.

Since the direct discrete sampling of the fGLE $\{V(t)\}_{t \in \bbN}$ does not generate an antipersistent process, for the purpose of comparison we opted for simulating the Cholesky and CME sequences based on the covariance function of the increment process  $Y_t = \nabla X(t) = X(t) - X(t-1)$, i.e.,
\begin{equation}\label{e:gamma_Delta_X}
\E {Y_t Y_{t+h}} = c^*(d)  \int_{\bbR} e^{i h \omega} \Big| \frac{1 - e^{-i \omega}}{i \omega} \Big|^2 \frac{1}{\nu_0 + \nu_1 |\omega|^{\beta} + \nu_2 |\omega|^{2 \beta}} \hspace{1mm}|\omega|^{2d} d\omega, \quad  t = 1,...,T,
\end{equation}
where $T = 2^9$, and $c^*(d) > 0$ is a constant. In turn, the wavelet simulation consisted of generating the fGLE, approximating $X(t) \approx \sum^{[(2^J/T)t]}_{k=1}V_{J,k} (T/2^J)$ and then sampling over $1, ..., T$. The wavelet filters are truncated at lag $|L| = 80$, which ensures that at the point of truncation the filters $u_j$ and $v_j$, $j = 1, ..., 13$, exhibit terms of magnitude below $10^{-5}$. The autocovariances (\ref{e:gamma_Delta_X}) and the integrals  of the initialization sequence (at $j = 0$) of the wavelet method were both calculated using quadrature-based numerical integration.  

The results are shown in Tables \ref{t:sim_fracGLE_dneg_length9_initj0procCME_compareJ} for $d = 0.10, 0.25, 0.45$. In all cases, the relative performance of the wavelet method seems to be fairly insensitive to the final approximation scale $J$. This indicates that it is not necessary to use a very fine scale in order to attain higher accuracy levels. On the other hand, since the wavelet method is the only approximate one among the three considered, it is \textit{a priori} expected to be more prone to affect the bias of the estimator. Indeed, the results seem to indicate that Cholesky and CME-based simulation tend to be slightly closer to each other, as measured by the estimator bias. Nevertheless, all three methods seem to be close enough, i.e., within two standard deviations.

\begin{table}[h]
\caption{fGLE: Local Whittle estimation of the parameter $d$ ($\gamma = 2$, $\textnormal{mass}=1$), comparison across values of $J$, $\Delta X(n)$, time series length $2^{9}$}
{\centerline {$u_j$, $v_j$ cut off at lag $|L| = 80$, init.\ CME time series length $2^{10}$  }}
\centering\begin{tabular}{ccccc}

$d = 0.10$ \\
\hline
method  & $\widehat{d}$ & $s$ & $N$ & $|t|$ statistic \\ \hline
wavelet (CME at  $j = 0$,  $J=6$) & $0.12044083$ &  0.09245007 & 5000 & 1.05431590\\
wavelet (CME at  $j = 0$,  $J=8$) & $0.12241433$ & 0.09102200 & 5000 & 0.02227624\\
wavelet (CME at  $j = 0$,  $J=10$) & $0.11932462$ & 0.09357999 & 5000 & 1.65282860\\
CME                         & $0.12221790$ & 0.09220211  &  5000  & 0.08515547\\
Cholesky & $0.12237381$ & 0.09088329  &  5000 & -\\
\hline\\
$d = 0.25$ \\
\hline
method  & $\widehat{d}$ & $s$ & $N$ & $|t|$ statistic \\ \hline
wavelet (CME at  $j = 0$,  $J=6$) & $0.27857299$ & 0.09217255  & 5000 & 0.25582876\\
wavelet (CME at  $j = 0$,  $J=8$) & $0.27998323$ & 0.09268102  & 5000 & 0.50777309\\
wavelet (CME at  $j = 0$,  $J=10$) & $0.28029380$ & 0.09627634 & 5000 & 0.66273074 \\
CME                         & $0.27818405$ &  0.09112314 &  5000  & 0.46947509\\
Cholesky  & $0.27904459$ &  0.09144735 &  5000 & -\\
\hline\\
$d = 0.45$ \\
\hline
method  & $\widehat{d}$ & $s$ & $N$ & $|t|$ statistic \\ \hline
AWD (patched filter, CME at  $j = 0$,  $J=6$) & $0.43156246$ & 0.10538368  & 5000 & 0.78943448\\
AWD (patched filter, CME at  $j = 0$,  $J=8$) & $0.43199573$ &  0.10660877 & 5000 & 0.58088823\\
AWD (patched filter, CME at  $j = 0$,  $J=10$) & $0.43166818$ & 0.10678505 & 5000 & 0.73448769\\
CME                         & $0.43338219$ & 0.10403330  &  5000  & 0.07274950\\
Cholesky  & $0.43322954$ &  0.10579033 &  5000 & -\\
\end{tabular}\label{t:sim_fracGLE_dneg_length9_initj0procCME_compareJ}
\end{table}

\newpage
%
%

\subsection{Challenges in Simulation.}


Simulation procedures are important to assist inferential efforts in multiple contexts in microrheological research.   Ideally, one is interested in fitting models based on memory kernels taken from a broad class of functions. It is still an open question whether CME and wavelet methods are well-suited in such a general context. This is of interest, for instance, in Monte Carlo-based quantification of sampling error.

Once one has fit a model for the diffusive properties of a material, there are a range of applications related to exploring the physical ramifications of the model.  In particular, there has been strong interest in characterizing the properties of human lung mucus to better understand the harmful effects of cystic fibrosis on lung health. One application of Monte Carlo methods is the study of first passage times to investigate drug delivery vehicles.  When a drug delivered via small particles is introduced into the lungs through such devices as inhalers, the particles will diffuse on the surface of the lung through mucus. By studying the first passage times of the particles, one gains a better understanding of the distribution of the particles throughout the internal lung surface.  A related challenge is to simulate diffusing particles in a properly modeled mucus layer, but with complex boundaries determined by the internal lung surface.  The particles will be reflected at this possibly non-smooth interface, a complex effect that needs to be taken into account in simulation efforts.

Two point microrheology expands the challenge of simulation to multiple coupled beads. In \cite{hohenegger2008two}, the focus is on the hydrodyamic interaction of two beads in close proximity. However, this effect does not have to be limited to only two. One of the reasons for using two is to simplify analytical results, and simulation would allow one to move beyond this limitation.  A related issue is to devise simulation methods for the full generalized Langevin equations including a possibly non-quadratic potential \eqref{eq:gle}.  The existence of solutions of the stochastic differential equation becomes much less clear; moreover, the typical methods for devising numerical schemes typically relies on an It\^o-type formula, which may not be readily available.


%

\section{Appendix}


The relation between the behavior of the spectral density at the origin and the decay of the autocorrelation function of a stationary process outside certain classes of parametric models is subject to specific assumptions (see \cite{gubner:2005}; for the antipersistent case, see \cite{bondon:palma:2007}). However, the related issue of the asymptotic behavior of the MSD can be addressed under general assumptions, as shown in the next proposition, used in Section \ref{s:subdiffusive}. The statement is restricted to anti-persistence (condition (\ref{e:behavior_f(x)_as_x->0})), but the same argument applies in the case of a singularity at the origin.\\

\noindent \textbf{Proposition}: \textit{Let $\{V(t)\}_{t \geq 0}$ be a zero mean Gaussian stationary (velocity) process with spectral density $\widehat{\rho}(\omega)$ satisfying
\begin{equation} \label{e:specdens_satisfies_Cramer_Leadbetter}
\int^{\infty}_{0} \omega^{2 \nu} \log(1+\omega) \widehat{\rho}(\omega) d\omega < \infty
\end{equation}
for some $\nu \in (0,1)$. Moreover, assume that $\widehat{\rho}(\omega)$ satisfies (\ref{e:behavior_f(x)_as_x->0}). Then (\ref{e:long_term_EX^2/t^alpha}) holds.}\\

\noindent {\bf Proof:}
Let
\begin{equation}\label{e:X=Riemann_V}
X(t) := \int^{t}_{0}V(s)ds
\end{equation}
where the integral (\ref{e:X=Riemann_V}) is taken in the Lebesgue sense. From Cram\'er and Leadbetter \cite{cramer:leadbetter:1967}, pp.\ 181-182, (\ref{e:X=Riemann_V}) is well-defined in the sense that there exists a process $\eta(t)$ which is equivalent to $V(t)$, and which satisfy a H\"{o}lder condition of order $\nu$ a.s. For notational simplicity, we will keep writing $V(t)$.

Let $\widehat{g}(x)$ be a spectral filter for the velocity process $V(t)$. Note that $s,s' \geq 0$,
$$
\E{|V(s)V(s')|} \leq \sqrt{\E{V^2(s)}}\sqrt{\E{V^2(s')}},
$$
which is finite and constant, by stationarity. Thus, based on the spectral integral representation of $V(t)$, by applying Fubini's Theorem and (\ref{e:X=Riemann_V}) to $\E{X(t)X(t')}$, $t,t'\geq 0$, we obtain the representation
\begin{equation}\label{e:specrepres_si_process}
X(t) \stackrel{{\mathcal L}}= \int_{\bbR} \Big(\frac{e^{it \omega}-1}{i \omega} \Big) \widehat{g}(\omega) d\widetilde{B}(\omega),
\end{equation}
where $\widetilde{B}(\omega) = \widetilde{B}_1(\omega) + i \widetilde{B}_2(\omega)$ is a complex-valued Brownian motion such that $d\widetilde{B}(-\omega) = \overline{d\widetilde{B}(\omega)}$ almost surely.

Condition (\ref{e:behavior_f(x)_as_x->0}) implies that
\begin{equation}\label{e:specdens_bound_above_below}
c_1 |\omega|^{2d} \leq \widehat{\rho}(\omega) \leq c_2 |\omega|^{2d}, \quad |\omega| \leq \delta,
\end{equation}
for some $c_1, c_2, \delta > 0$. Therefore,
\begin{equation}\label{e:EX^2}
\E{X^2(t)} = \int_{\bbR} \Big| \frac{e^{it \omega}-1}{i \omega}  \Big|^2 \widehat{\rho} (\omega)d\omega =  \Big( \int_{|\omega|\leq \delta} + \int_{|\omega| > \delta}  \Big) \Big| \frac{e^{it \omega}-1}{i \omega}  \Big|^2 \widehat{\rho}(\omega)d\omega.
\end{equation}
After a change of variables,
$$
\int^{\delta}_{-\delta} \Big| \frac{e^{it \omega}-1}{i \omega}  \Big|^2 |\omega|^{2d}d\omega \sim t^{1-2d} \int^{\infty}_{-\infty} \Big| \frac{e^{is}-1}{is}  \Big|^2 |s|^{2d}ds = t^{1-2d} \hspace{1mm} \E{ B^2_{1/2-d}(1)},
$$
where $B_{1/2-d}(\cdot)$ is a fBm with (Hurst) parameter $1/2-d$ (see Section \ref{sec:GLE}). Therefore, for large enough $t$, (\ref{e:EX^2}) can be bounded from above by
$$
c_2 t^{1-2d} \hspace{1mm} \E{ B^2_{1/2-d}(1)} + \int_{|\omega| > \delta}  \Big| \frac{2}{i\omega}\Big|^2 \widehat{\rho}(\omega) d\omega,
$$
where the second term is finite because $\widehat {\rho}(\omega) \in L^1(\bbR)$. On the other hand, for large enough $t$, (\ref{e:EX^2}) can be bounded from below by
$c_1' t^{1-2d} \hspace{1mm} \E{ B^2_{1/2-d}(1)}$ for some constant $c_1' > 0$. Therefore, (\ref{e:long_term_EX^2/t^alpha}) holds with the constants $c_1'$, $c_2$. $\Box$\\

The next proposition is used in Section \ref{s:subdiffusive}.\\

\noindent \textbf{Proposition}: \textit{Let $\{V(t)\}_{t \geq 0}$ be a stochastic process satisfying the assumptions of the previous Proposition. Additionally, assume that
\begin{equation} \label{e:rho_bounded_cont}
\textnormal{there exists a $0 < \delta < 2 \pi$ such that $\widehat{\rho}$ is bounded and continuous over $(-\delta,\delta)^c$}.
\end{equation}
Then (\ref{e:behavior_fY(x)_as_x->0}) holds.}\\

\noindent \textbf{ Proof:}
The discrete increments of (\ref{e:specrepres_si_process}) have the integral representation
$$
Y_j = \nabla X(j) = \int_{\bbR} e^{i j \omega} \Big( \frac{1 - e^{-i \omega}}{i \omega} \Big) \widehat{g}(\omega) \widetilde{B}(d \omega), \quad j \in \bbZ.
$$
For notational convenience, let
$$
\widehat{h}_{V}(\omega) = \Big| \frac{1 - e^{-i \omega}}{i \omega} \Big|^2 \widehat{\rho}(\omega).
$$
Then, by a standard calculation, the covariance of $Y_j$ is
$$
\rho_{Y}(j) = \int^{\pi}_{-\pi}e^{ij \omega} \sum_{k \in \bbZ} \widehat{h}_{V}(\omega + 2 \pi k) d\omega, \quad j \in \bbZ.
$$
Therefore, the spectral density of $Y$ can be expressed as
$$
\widehat{\rho}_{Y}(\omega) = \Big| \frac{1 - e^{- i \omega}}{i \omega}\Big|^2 \widehat{\rho}(\omega) + |1 - e^{-i \omega}|^2 \sum_{k \in \bbZ \backslash \{0\}} \widehat{\rho}(\omega + 2 \pi k) \frac{1}{|\omega + 2 \pi k|^2}, \quad \omega \in [-\pi,\pi).
$$
We are interested in the behavior of the ratio $\frac{\widehat{\rho}_{Y}(\omega)}{|\omega|^{2d}}$ as $\omega \rightarrow 0^{+}$. Note that, for small enough $\omega > 0$, $\omega + 2\pi k \in (-\delta,\delta)^c$, $k \in \bbZ \backslash\{0\}$. Therefore, by (\ref{e:rho_bounded_cont}) and the Dominated Convergence Theorem,
$$
\lim_{\omega \rightarrow 0^{+}} \sum_{k \in \bbZ \backslash \{0\}} \widehat{\rho}(\omega + 2 \pi k) \frac{1}{|\omega + 2 \pi k|^2} < \infty.
$$
The claim then follows from (\ref{e:behavior_f(x)_as_x->0}). $\Box$\\

\bibliographystyle{plain}
\bibliography{paper}

\end{document}